\documentclass[%
 reprint,
nofootinbib,
 amsmath,amssymb,
 aps,
floatfix,
]{revtex4-2}

\DeclareUnicodeCharacter{22C6}{\star}
\usepackage{graphicx}
\usepackage{dcolumn}
\usepackage{bm}

\usepackage{xcolor}
\usepackage{mathrsfs}

\begin{document}

\preprint{APS/123-QED}

\title{Sourced Carrollian Fluids Dual to Black Hole Horizons}

\author{Sercan Hüsnügil}
\email{shusnugil@perimeterinstitute.ca}
\affiliation{%
  Perimeter Institute for Theoretical Physics
}%
\affiliation{%
  University of Waterloo, Department of Physics and Astronomy
}%


\author{Luis Lehner}
 \email{llehner@perimeterinstitute.ca}
\affiliation{
 Perimeter Institute for Theoretical Physics
}%

\date{\today}

\begin{abstract}
The (degenerate) geometry of event horizons is linked to Carrollian fluids.  
We investigate the behavior of event horizons via a perturbative coupling to a massless scalar field, making connections to Carrollian hydrodynamics with a driving source, and discuss the fluid equilibration in tandem with the horizon's relaxation to equilibrium. We observe that after the perturbation dies off, the Carrollian fluid energy and momentum densities approach equilibrium as the horizon asymptotically becomes non-expanding. We connect the equilibration of the Carrollian fluid dual to the black hole horizon through the expansion of its background geometry.
\end{abstract}

\maketitle

\tableofcontents

\section{\label{secI:level1} Introduction}

Fully unraveling the behavior of black holes in dynamical settings is key to our understanding of gravity. They lie at the core of some of the most extreme (potentially) observable phenomena in our universe and challenge our understanding on many fronts. Of particular current interest is understanding their nonlinear behavior at a fundamental level, even when it is tractable via numerical simulations. That is, to extract robust, first-principles information beyond specific cases. 

The nonlinear behavior of field theories is notoriously difficult to explore. Perturbative approaches in such regimes become unreliable or require high-order computations, which might render them impractical. 
Exploiting possible analogies with hydrodynamics, a field that has traditionally had to embrace the nonlinear regime through multiple means, provides us with complementary techniques to explore them.
\textcolor{black}{In this work, we take this route to further this connection in gravity while also shedding light
on a particular set of hydrodynamical equations.}

The early work by Damour showed how the components of the Einstein equations, when projected on the black hole horizon, can be interpreted as the equation of energy conservation and the Navier--Stokes equations of a hypothetical horizon fluid~\cite{Damour1982}. These analogies were further developed within the membrane paradigm program~\cite{TPM1986, Price1986, Suen1985, Thorne1982}. There, the fluid analogy was constructed through projections on a timelike surface (the \textit{stretched horizon}) located outside the null event horizon. Such a timelike surface bypassed the singular nature of null surfaces and helped identify a fluid system with mechanical, electrical, and thermodynamic properties. The null nature of the event horizon obscured the structure of the resulting equations and was only recently interpreted as analogous to a special theory of fluids.

A closely related link between hydrodynamics and gravity was later pursued with important results. Motivated by AdS/CFT holography~\cite{Maldacena_1999}, a duality was derived between long-wavelength perturbations of
black holes in the bulk and relativistic hydrodynamics on the AdS boundary~\cite{Policastro_2002, Kovtun2003, Bhattacharyya_2008, hubeny2011fluidgravitycorrespondence}. Subsequent work connected phenomena with gravitational dynamics, first in asymptotically AdS~\cite{Carrasco_2012,Adams_2014,Green_2014, Raamsdonk_2008} and later in asymptotically flat spacetimes~\cite{Yang_2015,Iuliano:2024ogr,Ma:2025rnv}. Combined with the membrane paradigm, this duality has also been utilized to study transport coefficients in boundary theory, drawing insights from gravity \cite{Iqbal_2009}. 

In a more recent development, a duality of the properties of null surfaces and the vanishing speed of light, $c \rightarrow 0$ limit of relativistic hydrodynamics, named \textit{Carrollian fluids} was shown \cite{Donnay_2019}. This limit is opposite to the Galilean limit, where the speed of light is taken to be infinite. In both cases, the relativistic description, which puts temporal and spatial coordinates on equal footing, no longer holds: the Galilean limit results in an absolute time, whereas the Carrollian limit yields an absolute space \cite{Duval_2014}. 
Remarkably, the resulting equations—upon a suitable identification—can be linked to those obtained by the projection of Einstein equations onto a null surface; that are, Raychaudhuri and Damour equations.

Most of the work on the Carrollian fluid--gravity connection has so far focused on the kinematics of the vanishing-speed-of-light limit of relativistic theories and of null surfaces, in particular, the null infinity. While the notion of Carrollian fluids has also been studied as the limiting case of relativistic hydrodynamics, the Carrollian fluid dual to gravity has often been studied in vacuum spacetimes without a particular focus on the dynamics. Recently, Redondo-Yuste and Lehner~\cite{Redondo_Yuste_2023} took a dynamical point of view and studied the Carrollian fluid dual of a spherically symmetric black hole perturbed by gravitational waves. They interpreted their results to connect the properties of the fluid, such as its viscosity, to the metric perturbations by adapting the dictionary between Carrollian fluid parameters and the geometric properties of the horizon. 

In this work, we study the Carrollian fluid/gravity duality in the nonvacuum case. We follow the stretched Carrollian geometry formalism laid out in \cite{freidel2022carrollian,freidel2024geometrycarrollianstretchedhorizons} with specific coordinate choices while incorporating matter fields in the bulk spacetime. After establishing the general form of the equations, we focus on a massless scalar field in the bulk as an instructive example. We then concentrate on an ideal system with a spherically symmetric black hole coupled to an infalling massless scalar field and solve the coupled equations for fluid parameters perturbatively to demonstrate the effects of the existence of the matter fields. \textcolor{black}{We carry out perturbative computations up to second order in the scalar field strength as it is the lowest order at which the matter terms affect the horizon geometry.} Linking our results on the evolution of the horizon with the behavior of the dual Carrollian fluid, we connect the relaxation of the former to the equilibration of the latter. 

This work is organized as follows: In Section~\ref{secII:level1}, we lay out the geometric setup following the stretched Carrollian geometry formalism. We make judicial choices departing from the general formalism established in~\cite{freidel2024geometrycarrollianstretchedhorizons} to have a system of equations that is tractable with numerical methods. In Section~\ref{secIII:level1}, we summarize the results of previous work on constructing fluid dual to null surfaces, focusing specifically on the black hole horizon within our gauge choices. In Section~\ref{secIV:level1}, we focus on the deployment of sources in the bulk of spacetime and establish how these sources enforce changes in the gravitational equations describing the null surface. In Section~\ref{secV:level1}, we carry out a computation focusing on a Schwarzschild black hole perturbed by a scalar field to illustrate the effects of sources. We do so by employing a perturbative scheme and solving the resulting equations numerically. We conclude with a summary of our results and discussions in Section~\ref{secVI:level1}. 


\section{\label{secII:level1} Geometry: Horizon Embedding}
\textcolor{black}{The duality between the event horizon and fluids were naturally accommodated as a duality with Carrollian fluids~\cite{Donnay_2019, Adami_2021}. A dictionary has been established between gravitational (geometrical) parameters and fluid parameters~\cite{Donnay_2019, Redondo_Yuste_2023, freidel2022carrollian}. The recent work by Jai-akson and Freidel \cite{freidel2022carrollian, Freidel_2023, freidel2024geometrycarrollianstretchedhorizons} generalized the structure of Carroll symmetries and developed a framework that is applicable for both null and timelike hypersurfaces. They used this formalism to treat the problems of the membrane paradigm by defining a \textit{stretched} Carrollian structure on the timelike stretched horizons and obtaining the structure on the null horizon via taking a limit through the foliation of stretched horizons. }

In this section, we follow the stretched Carrollian structure \cite{freidel2022carrollian,freidel2024geometrycarrollianstretchedhorizons} which makes use of the null rigging formalism~\cite{Mars_1993}. The null rigging formalism allows a proper treatment of null surfaces embedded in an ambient spacetime, while the stretched Carrollian structure generalizes the notion of Carrollian symmetries to timelike hypersurfaces.
The combined formalism in \cite{freidel2022carrollian,freidel2024geometrycarrollianstretchedhorizons} allows us to construct a Carrollian structure induced from a bulk spacetime.\footnote{To study the Carrollian structure on the null surfaces, one does not need to introduce the stretching in practice. For example, see~\cite{Ciambelli:2023mvj, ciambelli2023nullraychaudhuricanonicalstructure}.}

Let $\mathscr{M}$ be a 4-dimensional spacetime manifold with signature $(-, +, + ,+)$,  which admits a null hypersurface $\mathscr{N}$.\footnote{The following treatment applies to Lorentzian manifolds of arbitrary dimensions. Here, we restrict ourselves to 4 dimensions for concreteness.} 
Let $\mathscr{H}_\rho$ for $\rho>0$ denote any member of a foliation of 3-dimensional timelike hypersurfaces (or \textit{stretched horizons}) embedded in $\mathscr{M}$ through an embedding function $r(x^a) = \rho$, where $\rho$ is called the ``stretching" \cite{freidel2024geometrycarrollianstretchedhorizons}. Here, $x^a$ denotes the coordinates on $\mathscr{M}$ and the letter $r$ is used for the foliation function to elucidate its radial characteristic. This foliation is constructed so that the null surface $\mathscr{N}$ is located at $r(x^a)=0$. Thus, taking the limit $r(x^a) = \rho \rightarrow 0$ corresponds to going from timelike hypersurfaces to the null surface in the ``center" of the radial foliation, \textit{i.e.,} $\mathscr{N} = \mathscr{H}_{\rho \rightarrow 0}$. In the following, we will use $\mathscr{H}$ dropping the subscript $\rho$ to denote a leaf of the foliation for which $r(x^a) = \rho$. 

A weak Carroll structure can be defined as a fiber bundle directly on the null hypersurface $\mathscr{N}$ if a null metric of the base manifold, $\mathcal{S}$, and an appropriate vector field is introduced on it \cite{Duval_2014_3, Ciambelli_2019, Freidel_2023}. When this vector field is interpreted as a time coordinate, the fiber bundle structure represents the splitting of time and space coordinates.\footnote{See also \cite{Duval_2014} for the relation between Carrollian structures to Newton-Cartan geometries.} 
On the timelike hypersurface $\mathscr{H}$, the stretched Carrollian structure can be defined by appropriate choices of a Carrollian vector field, the Ehresmann connection, the induced metric, and the stretching $\rho$.\footnote{Further details on the construction of Carroll structures can be found in the literature  \cite{Freidel_2023, Ciambelli_2019}.}  Next, we introduce these geometric ingredients for the Carrollian structure. 
\subsection{Vectors, Forms \& Projectors \label{secII:level2}}

Here, we summarize the recipe for the stretched Carrollian structure presented in \cite{freidel2024geometrycarrollianstretchedhorizons} for completeness. We then restrict ourselves to a particular gauge choice in the subsequent sections.

In the following, we reserve the lowercase Latin indices from the first half of the alphabet to denote the indices of tensors defined on $\mathscr{M}$. We use letters starting from $i$ for the intrinsic indices to the coordinates in $\mathscr{H}$. Finally, we use capital Latin letters to denote the coordinates intrinsic to the sphere manifold $\mathcal{S}$, which is the spacelike base manifold of the null hypersurface $\mathscr{N}$. Unless otherwise specified, we work with geometric units in which $c = G=1$.

We denote the metric on $\mathscr{M}$ by $g_{ab}$ and the covariant derivative compatible with it by $\nabla_a$. We also define on $\mathscr{M}$: a vector field $l^a$ that is tangent to $\mathscr{H}$, a normal form $n_a$, and a null \textit{rigging} vector $k^a$. 
These objects, along with their metric duals, satisfy the following relations:
\begin{align*}
     n_a k^a &= 1 \;,  &n_al^a= 0\;,   && k_a l^a &= 1 \;,\\
     n_a n^a &= 2\rho \;,  &k_a k^a = 0 \;, && l_a l^a&= -2\rho
\label{eq:one}.
\end{align*}
We further define a rigging projector $\Pi_a^{\hspace{4pt} b}$ in $\mathscr{M}$ as
\begin{equation}
    \Pi_a^{\hspace{4pt} b} \equiv \delta_a^b - n_a k^b \; ,
\end{equation}
which projects the tensors defined on $\mathscr{M}$ onto $\mathscr{H}$. Through this projector, we can relate the tangent vector field $l^a$ to the metric dual of the normal form, $n^a$, as $l^a = n^b \Pi_b^{\hspace{4pt}a}$. Therefore, we can decompose $n^a$ in terms of $l^a$ and $k^a$ as
\begin{equation}
    n^a =  l^a + 2\rho k^a. 
\end{equation}
This decomposition implies that the vectors $n^a$ and $l^a$ coincide on the null surface $\mathscr{N}$ since $\lim_{\rho \rightarrow 0} n^a = l^a$.

The induced metric on $\mathscr{H}$ is given by the projection of the bulk metric as
\begin{align}
      h_{ab} &=  \Pi_a^{\hspace{4pt}c} \, g_{cd} \, \Pi_b^{\hspace{4pt}d} \\
      &= g_{ab} - n_a k_b - n_b k_a \, .
 \end{align}
Restricting ourselves to the indices intrinsic to $\mathscr{H}$, we now have established the stretched Carrollian structure on $\mathscr{H}$ with the data $\left(l^i,  k_i, h_{ij}, \rho \right)$ \cite{freidel2024geometrycarrollianstretchedhorizons, freidel2022carrollian}.

The stretched Carrollian structure on the timelike hypersurface connects naturally to the Carrollian structure on the null hypersurface for the vanishing stretching $\rho$. To see this, we can decompose the metric on $\mathscr{H}$ as the following.
\begin{align}
    h_{ij} = q_{ij} - 2\rho k_i k_j \, .
\end{align}
Here, $q_{ij}$ is the null Carrollian metric that satisfies $l^i q_{ij}=0 $. Notice that $\lim_{\rho \rightarrow 0} h_{ij} = q_{ij}$. 

It is also useful to introduce the horizontal projector $q_a^{\hspace{4pt}b}$ which is defined as
\begin{equation}
    q_a^{\hspace{4pt}b} \equiv \Pi_a^{\hspace{4pt}b} - k_a l^b.
\end{equation}
This projector also projects the tensors on $\mathscr{M}$ to the horizontal base space of the bundle, $\mathcal{S}$. Using the horizontal projector, we can also induce the null Carrollian metric directly from the metric in the bulk, as
\begin{equation}
    q_{ab} = q_a^{\hspace{4pt}c} q_b^{\hspace{4pt}d} g_{cd} \, .
\end{equation}

The hypersurface $\mathscr{N}$ has the structure of a fiber bundle with a 2-dimensional base $\mathcal{S}$ and a one-dimensional vector field $l^a$ as its fiber \cite{Ciambelli_2019}. In this respect, the 1-form $k_a$ is also known as the Ehresmann connection, which allows the separation of vertical (along $l^a$) and horizontal directions in the fiber bundle \cite{freidel2022carrollian}. Through $k_a$, a basis frame $\left(e_A\right)^a$ is defined on the horizontal space, which satisfies $\left(e_A\right)^a k_a = 0$.  The basis of the horizontal subspace of $\mathscr{N}$ can be expressed with the frame $\left(e_A\right)^a$ and co-frame field $\left(e_A\right)_a$ satisfying
$q_{AB} = \left(e_A\right)_a \left(e_B\right)^a$. Here, the uppercase Latin indices are frame indices and are raised and lowered through $q_{AB}$, which is the metric on the sphere $\mathcal{S}$.   
 
To complement the fluid interpretation of the Carrollian structure, the following first-derivative quantities are introduced
\begin{align}
    \varphi_i &= -\mathcal{L}_l k_i \, , \\ 
    \varpi_{ij} &= -  q_i^{\hspace{4pt}k}  q_j^{\hspace{4pt}l} \partial_{\left[ \right.k} k_{\left.l\right]} \, , 
\end{align}
where the square brackets denote the anti-symmetrization of indices. The first quantity, $\varphi_i$, is the Carrollian acceleration; while the second quantity, $\varpi_{ij}$, is the Carrollian vorticity \cite{Donnay_2019}. These quantities also appear in the commutation relations of the vectors defined above, as demonstrated in \cite{freidel2022carrollian}.

Both the kinematics and the dynamics of the geometric structure introduced in this section can be conveniently studied by introducing two deviation tensors associated with the tangent vector $l^a$ and the rigging vector $k^a$. These tensors will prove useful as the deformation of the hypersurface $\mathscr{N}$ plays a crucial role in the dynamics of the dual Carollian fluid. We define the deviation tensors $\mathcal{S}$ as the following: 
\begin{align}
    B_{AB} & \equiv q_A^{\hspace{4pt}a} q_B^{\hspace{4pt}b} \nabla_b l_a \, , \\
    \bar{B}_{AB} & \equiv q_A^{\hspace{4pt}a} q_B^{\hspace{4pt}b} \nabla_b k_a \, . 
\end{align}
These deviation tensors can be decomposed into expansion (the trace), shear (the traceless symmetric part), and twist (the antisymmetric part), respectively, as follows:
\begin{align}
    B_{AB} &= \frac{1}{2} \Theta \, q_{AB} + \sigma_{AB} + \omega_{AB} \, ,\\[5pt]
    \bar{B}_{AB} &= \frac{1}{2} \bar{\Theta} \,  q_{AB} + \bar{\sigma}_{AB} + \bar{\omega}_{AB} \, .
\end{align}

Another useful information is the inaffinity, $\kappa$, of the tangent vector $l^a$. This term also has the interpretation of surface gravity for a black hole horizon. Inaffinity is given by the relation
\begin{align}
    l^b \nabla_b l^a = \kappa l^a \, .
\end{align}
Contracting with $k_a$, we obtain $\kappa = k_a l^b \nabla_b l^a $.

Finally, we define the Hájíček one-form 
\begin{equation*}
    \mathcal{H}_A \equiv q_A^{\hspace{4pt} b} k_c \nabla_b l^c \, .
\end{equation*}
This quantity is interpreted as an ``surface density of impulsion'' \cite{Damour:1979wya}, or fluid momentum by Damour~\cite{Damour1982}. Later, we will demonstrate the relation between the Hájíček one-form and the horizon angular momentum in a dynamical setting. 

The deviation tensors with their components, surface gravity, and the Hájíček field are the essential pieces for the fluid description of the horizon. As we shall discuss in the next sections, the stress-energy tensor describing $\mathscr{N}$, which admits the Carrollian fluid interpretation, is formulated using these geometric quantities. Moreover, the equations describing the dynamics of the horizon surface are conveniently described in terms of these quantities.

After the geometric description of the horizon parameters, we continue by introducing a set of coordinates that are convenient to illustrate the dynamics of the horizon and the effects of matter sources on it.

\subsection{The Metric \& Gauge Choices: Near Horizon\label{secII:level3}}

We focus on the event horizon of a black hole and the dual Carrollian fluid description of it. To work with an event horizon embedded in a 4-dimensional spacetime manifold $\mathscr{M}$, we adopt a version of the general form of the metric in \cite{freidel2024geometrycarrollianstretchedhorizons}.\footnote{Compared to the general metric given by Eq.$(62)$ of \cite{freidel2024geometrycarrollianstretchedhorizons}, we use a set of coordinates for which the coordinates restricted to the base manifold are the usual coordinates on a sphere, {i.e.,} $z^A = \sigma^A$, and the Jacobian matrix is given by a Kronecker delta, {i.e.,} 
 $J_A^{\hspace{4pt}B} = \delta_A^B$.} 
 
 On the other hand, to make contact with the near-horizon metric in Gaussian-null coordinates (as used in \cite{Donnay_2019,Adami_2021, Redondo_Yuste_2023}), we also set the Carrollian connection $b_A=0$ while allowing for a non-vanishing velocity field $V^A$. We further rescale the time coordinate to absorb the scale factor $\alpha$ so that it does not explicitly appear in our expressions.\footnote{Absorbing $\alpha$ into the time coordinate allowed the system of partial differential equations obtained through the perturbative scheme (section~\ref{secIII:level1} below) to close. An affine choice of $\alpha$ is further discussed in the Appendix~\ref{appendix:lgauge}.} With these choices, the metric on $\mathscr{M}$ is given by the line element\footnote{\textcolor{black}{Each constant-$r$ slice of this metric can be expressed in the Randers--Papapetrou parametrization \cite{Donnay_2019}.}}
\begin{align}\label{eq:metric}
    \mathrm{d}s^2_{\mathscr{M}} = \left( V^2- 2\rho \right) &\mathrm{d}v^2 + 2 \mathrm{d}v \, \mathrm{d}r  \notag \\ &- 2  V_A   \mathrm{d}v \, \mathrm{d} x^A + q_{AB} \mathrm{d}x^A \, \mathrm{d}x^B ; 
\end{align}
where $v$ denotes the null time, $r$ is the radial coordinate, and $x^A$ denotes the coordinates on a 2-sphere, and $V^2 \equiv q_{AB}V^AV^B$. Note that $x^a = \left(v, r, x^A\right)$ are the coordinates on $\mathscr{M}$, $x^i = \left(v, x^A\right)$ are the coordinates on $\mathscr{H}$ and $\mathscr{N}$, and $x^A$ are the coordinates on $\mathcal{S}$, which is the sphere base manifold of $\mathscr{N}$. As we will work with small perturbations of a Schwarzschild black hole in Section~\ref{secV:level1}, one can interpret the coordinates as perturbed versions of the advanced Eddington--Finkelstein coordinates.

After adopting the above gauge to obtain the metric, the stretching $\rho$, the horizontal velocity field $V^A$, and the sphere metric $q_{AB}$ are still functions of $x^a$ in general. Since we are interested in the dynamics of the horizon where $r = 0$, we focus on the near-horizon behavior and impose the following expansions\footnote{\textcolor{black}{Compare our choices with the Gaussian null coordinates used in Eq.(119-122) of \cite{freidel2024geometrycarrollianstretchedhorizons}.}}:
\begin{align}\label{eq:nearhorizonexp1}
    \rho(x^a) &= r \, \kappa(x^i)\\
     V^A(x^a) &= r \, U^A(x^i)  \label{eq:nearhorizonexp2}\\
     q_{AB}(x^a) &= \Omega_{AB}(x^i) - r \, \lambda_{AB}(x^i) \label{eq:nearhorizonexp3}
\end{align}
\textcolor{black}{Here, $\kappa$ is related to the surface gravity, $U^A$ is the horizontal vector field, $\Omega_{AB}$ is the sphere metric and $\lambda_{AB}$ are its first-order deformations in the radial coordinate. See Appendix \ref{appendix:examplegauge} for explicit examples of these functions in Schwarzschild and slowly-rotating Kerr spacetimes.} Thus, the metric we obtain agrees with those of \cite{Donnay_2019, Redondo_Yuste_2023} up to $\mathcal{O}\left(r^2\right)$. 
\begin{align}
    \mathrm{d}s^2_{\mathscr{M}} = &- 2 r  \kappa \, \mathrm{d}v^2 + 2 \mathrm{d}v \, \mathrm{d}r   \,- 2 r U_A \,  \mathrm{d}v \, \mathrm{d} x^A \notag \\ & \quad + \left(\Omega_{AB} - r \lambda_{AB}\right) \mathrm{d}x^A \, \mathrm{d}x^B + \mathcal{O}\left(r^2\right) 
\end{align}

\textcolor{black}{In the stretching formalism, each $r=\mathrm{const.}>0$ represents a timelike hypersurface. 
Note here that, after imposing the expansions in $r$, the limit $r\rightarrow 0$ (and $\rho\rightarrow0$) brings the timelike stretched horizons to the null event horizon.}
\begin{align}
 \mathrm{d}s^2_{\mathscr{N}} = 0 \, \mathrm{d}v^2 +  0 \,\mathrm{d}v \, \mathrm{d} x^A + \Omega_{AB} \,\mathrm{d}x^A \, \mathrm{d}x^B 
\end{align}
\textcolor{black}{We kept above the $v$ direction explicit for pedagogical reasons to emphasize that the event horizon is endowed with a 3-dimensional degenerate (null) metric.}

We express the tangent vector, the rigging vector, the normal form, and the horizontal frame in the above coordinate system as the following:
\begin{align*}
     l^a \partial_a &=   \partial_v  + V^A \partial_A , &k^a \partial_a &= \partial_r , \\
       n_a \mathrm{d}x^a &=  \mathrm{d}r\quad  &\left(e_A\right)^a \partial_a &=   \delta^B_A \partial_B.
\end{align*}
The metric duals of these are computed using the metric on $\mathscr{M}$:
\begin{align*}
      &l_a \mathrm{d}x^a = -2\rho \mathrm{d}v +  \mathrm{d}r , & k_a \mathrm{d}x^a &= \mathrm{d}v, \\
       &n^a \partial_a =  \partial_v + {2 \rho} \partial_r  + V^A \partial_A ,        & \hspace{-5pt}\left(e_{A}\right)_a \mathrm{d}x^a &= V_A \mathrm{d}v + q_{AB} \mathrm{d}x^B.
\end{align*}

With the near-horizon expansion enforced, we express the geometric quantities \textit{on the horizon} as the following:
\begin{align}
    \Theta  &\stackrel{\mathscr{N}}{=} \frac{1}{2} \Omega^{AB} \partial_v \Omega_{AB}\, ,\\[5pt]
    \sigma_{AB}  &\stackrel{\mathscr{N}}{=} \frac{1}{2} \partial_v \Omega_{AB} - \frac{1}{2} \Theta \Omega_{AB} \, ,\\[5pt] 
    \bar{\Theta} &\stackrel{\mathscr{N}}{=} \frac{1}{2} \Omega^{AB} \lambda_{AB}\, ,\\[5pt]
    \bar{\sigma}_{AB} &\stackrel{\mathscr{N}}{=} \frac{1}{2} \lambda_{AB} - \frac{1}{2} \bar{\Theta} \Omega_{AB} \, , \\[5pt]
    \mathcal{H}_A &\stackrel{\mathscr{N}}{=} \frac{1}{2} \Omega_{AB} U^B\, , 
\end{align}
where the twist tensors, $\omega_{AB}$ and $\bar{\omega}_{AB}$, vanish.
These geometric quantities will be used below to express the Einstein field equations projected on the horizon. Note that, throughout this text, we will use `` $\stackrel{\mathscr{N}}{=}$ " to denote the expressions which hold on the null horizon $\mathscr{N}$ but might or might not hold for a generic timelike stretched horizon $\mathscr{H}$, on which $\rho>0$.

With these expressions, the Carrollian acceleration ($\varphi_i = 0$) and vorticity ($\varpi_{ij} = 0$) should vanish on any $\mathscr{H}$. Furthermore, vanishing $b_A$ implies that the Levi-Civita--Carroll covariant derivative simply agrees with the covariant derivative induced on the 2-sphere, $\hat{\nabla}_A \equiv \nabla_A$. These simplifications make the analysis of dynamical equations considerably simpler. See Appendix~\ref{appendix:derivatives} for more details on the Carrollian and rigged derivatives.

Note that the gauge choices and conventions we use here simplify the computations on the black hole horizon drastically. In Appendix~\ref{appendix:bgauge}, implications of different gauge choices regarding $b_A$ and $V^A$ in the context of horizon dynamics are explored. Also, in Appendix~\ref{appendix:examplegauge}, the coordinates and the geometric quantities are given explicitly for Schwarzschild and slowly-rotating Kerr spacetimes. 

\section{\label{secIII:level1} Horizon Dynamics \& Fluid Interpretation}

\textcolor{black}{
The study of the Carrollian limit began with the work of Lévy-Leblond~\cite{Levyleblond1965} and Gupta~\cite{Gupta1966} as a mathematical exploration of the contraction~\cite{Inonu1953} of the Poincaré group, in which the speed of light vanishes. These studies have evolved into an active field in physics in the last decade~\cite{LevyLeblond2022}. Some of the works on the field focused on the relations between the asymptotic symmetries of asymptotically flat spacetimes with Carroll groups and their use in flat holography~\cite{Duval_2014_2,Ciambelli_2018_2, Ciambelli_2019,Ciambelli_2019_2, ostendo:fontanella2025carrolllimitadscfttriality}; Carrollian \textit{fluids} as the limit of relativistic hydrodynamics~\cite{Ciambelli_2018, Ciambelli2019,Petkou_2022,Freidel_2023}; the interacting particle motion~\cite{Bergshoeff_2014} and dynamics of extended objects~\cite{Cardona_2016}; Carrollian limits of relativistic field theories~\cite{Basu_2018, Henneaux_2021, Gupta_2021, ostendo:Hao_2022, Hansen_2022, ostendo:Bagchi_2023, ciambelli2023carrollgeodesics,Ecker_2023, Baiguera_2023, ciambelli2023dynamicscarrollianscalarfields}\footnote{Note also the early work (1998) by Dautcourt~\cite{dautcourt1998ultrarelativisticlimitgeneralrelativity} on the ``ultrarelativistic'' limit of general relativity \textcolor{black}{and the book (2000) by Klauder \cite{Klauder:2000ud} for ultralocal models in quantum field theories.}}; applications in cosmology with emphasis on dark energy \cite{deboer2021carrollsymmetrydarkenergy, deboer2023carrollstories}.} 

\textcolor{black}{
Thermodynamic properties of Carrollian fluids have also been studied in recent works: for example, in \cite{ostendo:Armas_2024} from the point of view of symmetries (as was developed in a general framework in \cite{ostendo:de_Boer_2018}), and in \cite{ostendo:arenashenriquez2025radiation} within a holographic setup by relating the gravitational radiation with the dissipation in the dual Carrollian theory on the boundary. Despite the progress in the field, Carrollian quantum field theories are not yet fully understood as challenges regarding ill-defined partition functions \cite{deboer2021carrollsymmetrydarkenergy, deboer2023carrollstories}, gauge dependence of mass and coupling parameters \cite{Mehra_2023,sharma2025studiescarrollianquantumfield}, sensitivity to UV and IR regimes as well as exhibiting a UV/IR mixing \cite{Cotler_2024} remain.} 

\textcolor{black}{In this section, we discuss the equations that describe the dynamics of the horizon and their interpretation as a hydrodynamical description of a dual Carrollian fluid.}

The governing equations of gravitational dynamics in the spacetime are the Einstein field equations. These equations, when projected onto the hypersurface $\mathscr{H}$, can be regarded as the conservation equations of a stress-energy tensor defined on $\mathscr{H}$. This stress-energy tensor, informed by the extrinsic geometry of the hypersurface $\mathscr{H}$, is constructed via
\begin{equation}
    T_i^{\hspace{4pt}j} = N_i^{\hspace{4pt}j} - N \Pi_i^{\hspace{4pt}j},
    \label{eq:GEOstressenergy}
\end{equation}
where $N_i^{\hspace{4pt}j}$ is the generalized news tensor and $N$ is its trace \cite{freidel2024geometrycarrollianstretchedhorizons}.\footnote{See \cite{freidel2024geometrycarrollianstretchedhorizons} and \cite{Riello_2024} for the definition of the generalized news tensor. Here, we omit a detailed discussion as our goal is to focus on the null hypersurface.} On the null surface $\mathscr{N}$, this tensor becomes the Weingarten map of the surface, i.e., $\lim_{\rho\rightarrow0} N_i^{\hspace{4pt}j}= W_i^{\hspace{4pt}j}$, which is defined by $ W_i^{\hspace{4pt}j} \equiv  \Pi_i^{\hspace{4pt}c} \left( \nabla_c n^d \right)  \Pi_d^{\hspace{4pt}j}$. This definition is compatible with the limit $\rho \rightarrow 0$ and hence is also valid on the null surface $\mathscr{N}$ \cite{Chandrasekaran_2022, freidel2022carrollian}. For the gauge we consider in this paper, the generalized news tensor and the Weingarten map also agree on any $\mathscr{H}$. 

The projection of the Einstein tensor $G_{ab}$ on $\mathscr{M}$ along $n^a$ can be described by the conservation of the stress-energy tensor on $\mathscr{H}$ \cite{freidel2022carrollian} 
\begin{equation}
    \Pi_i^{\hspace{4pt}b} n^c G_{cb} = D_j T_i^{\hspace{4pt}j}.
\end{equation}
Here $ D_i T_j^{\hspace{4pt}k} \equiv \Pi_i^{\hspace{4pt}a} \nabla_a T_j^{\hspace{4pt}k}$ is the rigged derivative of the stress-energy-tensor on $\mathscr{H}$.
Through the realization of the vacuum Einstein field equations on $\mathscr{M}$, this equation admits a conservation equation for the stress-energy tensor on $\mathscr{H}$:
\begin{equation}
    G_{ab} = 0 \Rightarrow  D_j T_i^{\hspace{4pt}j} = 0.
\label{eq:vacuum}
\end{equation}
On the timelike hypersurface $\mathscr{H}$, the stress-energy tensor can be expressed as the energy-momentum tensor of a fluid field defined on $\mathscr{H}$. In that case, the Einstein equations in vacuum ensure the fluid stress-energy tensor is conserved; thus, the equations of relativistic fluid dynamics are satisfied as expressed in Eq.~(\ref{eq:vacuum}). 

On the null surface $\mathscr{N}$, Eq.~(\ref{eq:vacuum}) is still satisfied in vacuum. However, the stress-energy tensor in this case has to be interpreted as a Carrollian stress-energy tensor. 
The resulting equations are not the equations of relativistic hydrodynamics but those of Carrollian hydrodynamics \cite{Ciambelli_2018, freidel2022carrollian, freidel2024geometrycarrollianstretchedhorizons}. Therefore, going from the timelike $\mathscr{H}$ to null $\mathscr{N}$ via the limit $\rho \rightarrow 0$ intuitively corresponds to taking the $c \rightarrow 0$ limit of the relativistic hydrodynamics equations to obtain the Carrollian ones \cite{Donnay_2019}.

\subsection{Carrollian Fluid/Horizon Dictionary \label{secIII:level2}}

The particular case of Carrolllian fluid dynamics and its dual interpretation as horizon dynamics can be drawn by comparing the Carrollian fluid equations of motion~\cite{Ciambelli_2018, Ciambelli2019} with geometrical (gravitational) equations describing the motion of the null surface in the ambient spacetime.  These geometrical equations are commonly referred to as the null Raychaudhuri and Damour--Navier--Stokes (DNS) equations \cite{Donnay_2019, Adami_2021}. Below, we express these equations \textit{on the horizon} in the respective order, imposing the gauge above:
\begin{align}
\hspace*{-1cm}
   -G_{ll} &\stackrel{\mathscr{N}}{=} \partial_v {\Theta} - \kappa \Theta + \frac{1}{2} \Theta^2 + \sigma_{AB} \sigma^{AB} \, ,  \label{eq:Raychaudhuri} \\[5pt]
    G_{lA} &\stackrel{\mathscr{N}}{=} \partial_v{\mathcal{H}}_A  +{\Theta} \mathcal{H}_A  + {\nabla}_B \sigma_A^{\hspace{4pt}B} - {\nabla}_A \left(\kappa +\frac{\Theta}{2}\right) \label{eq:DNS}   \, . 
\end{align}
Here, the subscripts of the Einstein tensor denote the contractions with the relevant vectors, that is, $G_{ll} \equiv l^a l^b G_{ab} $ and $G_{lA} \equiv l^a \left(e_A\right)^b G_{ab} $. In vacuum, the right-hand sides of both of these equations should evaluate to $0$ as the Einstein tensor vanishes.

The fluid interpretation of the horizon geometry also follows from the terms appearing in the stress-energy tensor. As constructed using the Weingarten map, the stress-energy tensor defined on $\mathscr{H}$ is given by~\cite{freidel2024geometrycarrollianstretchedhorizons}:
\begin{align}
    T_i^{\hspace{4pt}j} = -k_i \left(\mathcal{E} l^j + \mathcal{J}^j  \right)+ \mathcal{P}  q_i^{\hspace{4pt}j} + \pi_i l^j +\tau_i^{\hspace{4pt}j}
\end{align}
Here, the scalars $\mathcal{E}$ and $\mathcal{P}$ are the energy density and the pressure of the fluid, respectively. The one-form $\pi_i$ is interpreted as the fluid momentum density, whereas the vector $\mathcal{J}^i$ is the heat flux. The tensor $\tau_i^{\hspace{4pt}j}$ is identified with the viscous stress tensor of the fluid. Evaluating this stress-energy tensor for the black hole horizon~\cite{freidel2024geometrycarrollianstretchedhorizons}
The following Carrollian fluid/horizon geometry identifications are made \textit{on the horizon:}
\begin{align*}
    \mathcal{E} &\stackrel{\mathscr{N}}{=} \Theta \, ,  &\quad&  \mathcal{P} \stackrel{\mathscr{N}}{=} - \left( \kappa + \frac{\Theta}{2} \right) \,,  \\[5pt]
    \tau_{AB} &\stackrel{\mathscr{N}}{=} 2 \eta \, \sigma_{AB} \, , &\quad&  \pi_A \stackrel{\mathscr{N}}{=} \mathcal{H}_A \;. 
\end{align*}
while $\mathcal{J}^A \stackrel{\mathscr{N}}{=} 0$. The shear viscosity is given by $\eta = 1/2$. Note that $\mathcal{P}$ encodes the pressure of the Carrollian fluid \textit{and} the ambient pressure in the sense that $\mathcal{P}$ is non-zero even when the energy density of the fluid vanishes. It i the surface gravity $\kappa$ that plays the role of the ambient pressure for the Carrollian fluid through this dictionary identification. This definition of pressure agrees with the one in \cite{freidel2022carrollian}. {\em This dictionary identification need not be modified when the matter fields are introduced in the bulk of the spacetime.}

To have a closed system of equations that can be solved consistently, we need two more equations obtained from different projections of the Einstein tensor (e.g.~\cite{G_mez_2001,Redondo_Yuste_2023}). We choose these components to be the traceless symmetric part of the horizontal projection of the Einstein tensor and the mixed tangential-radial projection, which are expressed \textit{on the horizon} with our gauge choices as follows:
\begin{align}
     -G_{\langle AB \rangle} &\stackrel{\mathscr{N}}{=}  2\partial_v{\bar{\sigma}}_{AB} - 2 \Omega_{AB}\left(\bar{\sigma}_{CD} \sigma^{CD} \right) + \left( 2\kappa - \Theta \right) \bar{{\sigma}}_{AB} \notag \\[5pt]
     &  \hspace{-5pt}+ \bar{\Theta}\sigma_{AB} 
    +2 {\nabla}_{\langle A} \mathcal{H}_{B\rangle} + 2  \mathcal{H}_{\langle A}  \mathcal{H}_{B\rangle} - \hat{R}_{\langle AB \rangle}
\end{align}
\begin{align}
     G_{lk} \stackrel{\mathscr{N}}{=} \partial_v{\bar{\Theta}} + \left(\kappa + \Theta \right) \bar{\Theta} + {\nabla}_A \mathcal{H}^A + \mathcal{H}_A \mathcal{H}^A - \frac{1}{2} \hat{R} 
\end{align}
where $\hat{R}_{AB}$ is the Ricci tensor defined with respect to the metric $q_{AB}$ using Levi-Civita--Carroll covariant derivatives. Note that $G_{\langle AB \rangle} \equiv \left(e_{\langle A}\right)^a \left(e_{B \rangle}\right)^b G_{ab}$ and  $G_{lk} \equiv l^a k^b G_{ab}$, where angular brackets denote the symmetric traceless part of the tensors. For brevity, we will refer to the system of four coupled partial differential equations introduced in this section as the { \textit{horizon equations}} throughout this text.

\section{\label{secIV:level1} Fluid Sources} 

We introduce matter fields in the bulk spacetime and examine the properties of the Carrollian fluid dual to the horizon. We start from the sourced equations of the relativistic fluid dual to $\mathscr{H}$ and take the limit $\rho \rightarrow 0$ to recover the Carrollian equations for the fluid on $\mathscr{N}$.

In the presence of a non-vacuum stress-energy tensor $\mathcal{T}_{ab}$ in $\mathscr{M}$, the Einstein field equations imply the following relation for the fluid stress-energy tensor on $\mathscr{H}$:
\begin{equation}
    G_{ab} = 8 \pi \mathcal{T}_{ab} \Rightarrow  D_j T_i^{\hspace{4pt}j} = \Pi_i^{\hspace{4pt}b} n^c \mathcal{T}_{cb} \, ,
    \label{eq:sourced}
\end{equation}
where we set the natural constants to 1.

Schematically, we would like to relate the reaction of the Carrollian fluid to the stress-energy tensor on $\mathscr{M}$ via
\begin{equation}
    \lim_{\rho \rightarrow 0} D_j T_i^{\hspace{4pt}j} = S_i \, ,
    \label{eq:sourcelimit}
\end{equation}
where $S_i$ describes a generic source field defined on $\mathscr{N}$. 
Thus, we can study the behavior of the Carrollian fluid dual to the horizon in a spacetime with non-vacuum sources via modeling the fluid equations subject to sources and sinks.

We continue by computing the corresponding source fields for an example matter field on $\mathscr{M}$. Consider a massless scalar field in $\mathscr{M}$ described by the following stress-energy tensor: 
\begin{equation}
   \mathcal{T}_{ab} = \nabla_a \Phi \nabla_b \Phi - \frac{1}{2} g_{ab} \left (g^{cd} \nabla_c \Phi \nabla_d \Phi \right)  \; .
\end{equation}
Here, the scalar field is defined on $\mathscr{M}$, that is, $\Phi = \Phi (x^a)$.  However, near the horizon, we will assume that $\Phi = \Phi (x^i)$ as the field should depend on the radial coordinate through the null time $v$ \cite{Iqbal_2009}.
Plugging this stress-energy tensor in  Eq.~(\ref{eq:sourced}), we obtain the following source term on $\mathscr{N}$:
\begin{equation}
    S_i = l^j D_j \Phi D_i \Phi \, .
\end{equation}

Explicitly, the existence of the scalar field in the bulk invokes the following contributions to the horizon equations through the field equations:\footnote{\textcolor{black}{Observe that for a generic $\Phi\left(v,r,\theta,\phi\right)$, the change of the radial profile of the scalar field, $\partial_r\Phi$, plays a role in the last two equations even on the horizon where the stretching vanishes. As expressed before, we will neglect the contributions from $\partial_r \Phi$ near the horizon, which will further simplify our treatment.}}
\begin{align}
G_{ll} &\stackrel{\mathscr{N}}{=}   8 \pi\left(\partial_v \Phi \right)^2 \\[5 pt]
G_{lA} &\stackrel{\mathscr{N}}{=} 8 \pi \partial_v \Phi {\partial}_A \Phi \\[5 pt]
G_{\langle AB \rangle} &\stackrel{\mathscr{N}}{=}  8 \pi \left[{\partial}_A \Phi {\partial}_B \Phi \right. \notag \\[5pt]
&\quad \left.-\frac{1}{2} \Omega_{AB} \left( \Omega^{CD} {\partial}_C \Phi {\partial}_D \Phi + 2\partial_v \Phi \partial_r \Phi \right) \right]  \\[5 pt]
G_{kl} &\stackrel{\mathscr{N}}{=} - 4 \pi \left(\Omega^{AB} {\partial}_A \Phi {\partial}_B \Phi  + 2\partial_v \Phi \partial_r \Phi \right)
\end{align}

Next, we will study the horizon equations with sources in a perturbative expansion on the scalar field amplitude and solve the system numerically.

\section{\label{secV:level1} Perturbative Treatment} 

Nonlinear effects have recently garnered renewed attention in black hole perturbation theory, with a focus on analyzing the ringdown phase of gravitational wave signals. Some of these efforts study the imprints of nonlinearities induced by the non-stationary nature of the black hole as it accretes energy and angular momentum (effectively resulting in changes to the black hole parameters) on the ringdown signal \cite{Sberna_2022, May_2024}. Others make use of second-order black hole perturbation theory to study mode couplings and the generation of quadratic quasinormal modes during ringdown~\cite{Loutrel_2021, Ripley_2021,Ma_2024, Bucciotti_2024, Khera_2023, Khera_2025}. A recurring theme is the presence of a transient phase whose complexity exceeds that of a simple decomposition into linear modes. This highlights the need for novel approaches to capture the dynamics of this stage and to elucidate the mechanisms governing the transition toward linear behavior on an 
effectively static background (e.g.~\cite{Yang_2015,Iuliano:2024ogr,Ma:2025rnv}). 

Here, we focus on a black hole perturbed by a surrounding matter field and perturbatively solve for the back-reaction on the horizon geometry. In the process, we also connect with the question of driving Carrollian fluids and their equilibration.

Black holes interact with their surroundings. 
The gravitational field of the matter in the bulk affects the dynamics of the horizon. Furthermore, the infalling matter to the black hole results in the expansion and shearing of the horizon.
We study such a dynamical scenario by adopting a perturbative scheme. We focus on a Schwarzschild black hole perturbed by an infalling massless scalar field through a small coupling parameter, $\epsilon$. We follow a numerical implementation similar to that in \cite{Redondo_Yuste_2023}, while incorporating matter fields.

We expand the full spacetime metric as
\begin{equation}
    g_{ab} = g_{ab}^{(0)} + \epsilon g_{ab}^{(1)} + \epsilon^2 g_{ab}^{(2)} \;,
\end{equation}
where $g_{ab}^{(0)}$ is the Schwarzschild metric. We can write this metric  in the vicinity of the horizon, in a form similar to that of Eq.~(\ref{eq:metric}) as the following:
\begin{align}
    g_{ab}^{(0)} = - \frac{\tilde{r}}{2M} \mathrm{d}v^2 &+ 2 \mathrm{d}v \, \mathrm{d}\tilde{r} \notag \\
    &+ 4M\left(M + \tilde{r}  \right) \left(\mathrm{d} \theta^2 + \sin{\theta}^2 \mathrm{d} \phi^2  \right) .
\end{align}
Here, $\tilde{r} \equiv 2M\left(1 - {r_S}/{r}\right)$ is a redefinition of Schwarzschild radial coordinate $r$, for which the horizon is located at  $\tilde{r} =0$, and $r_S = 2M$ is the Schwarzschild radius in geometrized units. This redefinition facilitates the near-horizon expansion. We drop the tilde over $\tilde{r}$ and treat this new coordinate as the radial coordinate in the following expressions. Comparing this metric with  Eq.~(\ref{eq:metric}) we identify the terms of order $\mathcal{O}(1)$ (zeroth order in $\epsilon$):
\begin{align*}
   \rho^{(0)} = \frac{r}{4M} \, , && 
   V_A^{(0)} = 0 \, ,  && q_{AB}^{(0)} = 4M (M +r) \Omega_{AB}^{(\text{Sph})} \, ,
\end{align*}
where $\Omega_{AB}^{(\text{Sph})}$ denotes the metric components for a unit 2-sphere. As we work in the vicinity of the horizon,  through Eq.~(\ref{eq:nearhorizonexp1}), these relations imply
\begin{align*}
   \kappa^{(0)} &= \frac{1}{4M} \, , && 
   U_A^{(0)} = 0 \, ,  && \\ \Omega_{AB}^{(0)} &= 4M^2 \Omega_{AB}^{(\text{Sph})} \, , && \lambda_{AB}^{(0)} = 4M \Omega_{AB}^{(\text{Sph})} \, .
\end{align*}

We further assume the trace of the higher-order perturbations in the sphere metric is $0$. That is, 
\begin{equation}
\Omega^{\text{(Sph)}AB}q_{AB}^{(1)}=\Omega^{\text{(Sph)}AB}q_{AB}^{(2)} =0 \, .
\end{equation}

In the following, we write all the functions in the line element up to second order in $\epsilon$:
\begin{align*}
   \rho\left( r\right) &= \frac{r}{4M} + \epsilon \rho^{(1)} + \epsilon^2 \rho^{(2)}\; ,\\[5pt] 
   V_A\left( v, \theta, \phi\right) &=  0 + \epsilon V_A^{(1)} + \epsilon^2 V_A^{(2)}\; ,\\[5pt]
   q_{AB}\left(v, r, \theta, \phi\right) &= 4M (M + r ) \Omega_{AB}^{(\text{Sph})} + \epsilon q_{AB}^{(1)} + \epsilon^2 q_{AB}^{(2)} \;. 
\end{align*}
These expansions are kept up to 
$\mathcal{O}\left(\epsilon^2\right)$, since this is the order in which the contributions of the scalar field in the equations of motion appear.

We further prescribe an expansion for the infalling scalar field in the vicinity of the horizon: 
\begin{align}
    \Phi\left( v,\theta, \phi\right) = \epsilon \Phi^{(1)} + \epsilon^2  \Phi^{(2)} \, ,
\end{align}
which is also up to $\mathcal{O}\left(\epsilon^2\right)$. Note that we assume here the scalar field has a constant radial profile near the horizon.

Using the expansions of the metric parameters above, we expand the geometric quantities \textit{on the horizon} in orders of the coupling $\epsilon$:
\begin{align}   
    \kappa &\stackrel{\mathscr{N}}{=} \frac{1}{4M} + \epsilon \kappa^{(1)} + \epsilon^2 \kappa^{(2)}\;,  \\[5pt] 
    \mathcal{H}_A &\stackrel{\mathscr{N}}{=} -\frac{1}{2} \epsilon U_A^{(1)} -\frac{1}{2} \epsilon^2 U_A^{(2)}\;,  \\[5pt] 
    \Theta &\stackrel{\mathscr{N}}{=}  -\epsilon^2 \frac{1}{2} \Omega^{AC}_{\text{(Sph)}}\Omega^{BD}_{\text{(Sph)}}\Omega_{CD}^{(1)}\partial_v \Omega_{AB}^{(1)}\;,  \\[5pt] 
     \bar{\Theta} &\stackrel{\mathscr{N}}{=} \frac{1}{M} + \frac{1}{2M}\epsilon^2 \Omega^{AB(1)} \left( \Omega_{AB}^{(1)} - M \lambda_{AB}^{(1)}\right) \;,\\[5pt]
     \sigma_{AB} &\stackrel{\mathscr{N}}{=} \frac{1}{2} \epsilon \partial_v \Omega_{AB}^{(1)} \notag \\
     &+ \frac{1}{2} \epsilon^2 \left[\partial_v \Omega_{AB}^{(2)} +  \left(\Omega^{CD(1)} \partial_v \Omega_{CD}^{(1)} \right) \Omega^{\text{(Sph)}}_{AB}\right]\;, 
\end{align}
\begin{align}
   \bar{\sigma}_{AB} &\stackrel{\mathscr{N}}{=}  \epsilon \frac{1}{2M} \left(M \lambda_{AB}^{(1)} - \Omega_{AB}^{(1)}  \right) + \epsilon^2 \frac{1}{2M}\left[ M  \lambda_{AB}^{(2)}  \right. 
    \notag\\
    &  - \Omega_{AB}^{(2)}+ \left. \Omega^{CD(1)} \left( M  \lambda_{CD}^{(1)} -\Omega_{CD}^{(1)}\right) \Omega^{\text{(Sph)}}_{AB} \right]\;.
\end{align}

The source terms induced from the scalar field in the bulk onto the horizon all appear in $\mathcal{O}\left(\epsilon^2\right)$ and are given by the following \textit{on the horizon}:
\begin{align}
\mathcal{T}_{ll} &\stackrel{\mathscr{N}}{=} \epsilon^2\left(\partial_v \Phi^{(1)} \right)^2 \\[5 pt]
\mathcal{T}_{lA} &\stackrel{\mathscr{N}}{=} \epsilon^2\partial_v \Phi^{(1)} {\partial}_A \Phi^{(1)} \\[5 pt]
\mathcal{T}_{\langle AB \rangle} &\stackrel{\mathscr{N}}{=} \epsilon^2\left[ {\partial}_A \Phi^{(1)} {\partial}_B \Phi^{(1)} - \right.  \notag \\ 
& \qquad \left. \frac{1}{2} \Omega_{AB}^{(\text{Sph})} \left( \Omega^{(\text{Sph})CD} {\partial}_C \Phi^{(1)} {\partial}_D \Phi^{(1)} \right)  \right]  \\[5 pt]
\mathcal{T}_{kl} &\stackrel{\mathscr{N}}{=} -\epsilon^2\frac{1}{2} \Omega^{AB} {\partial}_A \Phi^{(1)} {\partial}_B \Phi^{(1)} 
\end{align}

Plugging in these expansions to the projections of the Einstein tensor onto the horizon $\mathscr{N}$, we obtain a set of equations up to $\mathcal{O}(\epsilon^2)$ given in Appendix \ref{appendix:equationsO2}.

These equations can be re-expressed as a first-order coupled system of partial differential equations for the fields $\{q_{AB}, \mathcal{H}_A, \kappa , \Theta \}$ in the orders of $\epsilon$ and $\epsilon^2$. The resulting system can be solved order by order in $\epsilon$ given the information from the bulk $\{ \lambda (=\partial_r q_{AB}), \Phi\}$ in the respective orders. Note that both $\hat{R}$ and $\Theta$ can be expressed solely in terms of $q_{AB}$; however, we include them in the equations for the ease of notation.

\textcolor{black}{We shall solve the horizon equations (given explicitly in Appendix \ref{appendix:equationsO2}) numerically. Because the quantities we work with are defined on the sphere, we introduce spin-weighted functions and recast the equations in terms of these.}

\subsection{Spin-weighted Functions \label{secV:level2}}
To solve the set of coupled partial differential equations, we follow the spin-weighted formalism \cite{Goldberg:1966uu,Gomez_1997, Bishop_1997}. We do so by defining the complex dyad $q^A$ which satisfies the following relations:
\begin{align}
    q^A q_A &= \bar{q}^A \bar{q}_A = 0 \, , \\
    q^A \bar{q}_A &= \bar{q}^A q_A = 2 \, , \\
    \Omega^{\text{(Sph)}}_{AB} &= q_{\left(A \right.} \bar{q}_{\left. B \right)} \, .\label{metric-to-dyad}
\end{align}

We then express the vector and tensor quantities in terms of spin-weighted scalar coefficients by contracting the quantities with the dyad as follows: 
\begin{align}
    \Omega_{AB}^{(1)} &= \frac{1}{2}\left(c \bar{q}_A \bar{q}_B  + \bar{c} q_A q_B \right) \, , \\
    \lambda_{AB}^{(1)} &= \frac{1}{2}\left(s \bar{q}_A \bar{q}_B  + \bar{s} q_A q_B \right) \, , \\
    \mathcal{H}_A^{(1)} &= \frac{1}{2} \left(h \bar{q}_A + \bar{h} q_A \right) \, .
\end{align}
Similar relations can be written for terms of order $\mathcal{O}(\epsilon^2)$, for which we use $\{C,S,H\}$ as the respective spin-weighted functions.

Note that the perturbations on the induced metric $\Omega_{AB} $ and $\lambda_{AB}$ are assumed to be traceless. Therefore, $\bar{q}^Aq^B\Omega_{AB} = 0 = \bar{q}^Aq^B\lambda_{AB}$. We also rename the scalar quantities to follow the convention in \cite{Redondo_Yuste_2023} and for the ease of notation:
\begin{align}
     \kappa^{(1)} &= k, \qquad \kappa^{(2)} = K, \\
    \Theta^{(2)} &= \vartheta, \qquad \Phi^{(1)} = \varphi \, .
\end{align}

Following \cite{Gomez_1997}, we compute the curvature scalars on the horizon in terms of the spin-weighted scalars:
\begin{align}
        \hat{R}^{(1)} &=\frac{1}{32M^4} \left(\bar{\eth}^2c + \eth^2 \bar{c} \right) \, ,\\[5pt]
        \hat{R}^{(2)} &= \frac{1}{32M^4} \left[\bar{\eth}^2 C + \eth^2 \bar{C}  \notag \right. \\ & \qquad \qquad \quad \left.+ \frac{1}{8} \left(\eth c \bar{\eth} \bar{c} - \eth \bar{c} \bar{\eth} c + 8 c\bar{c} \right) \right] \, .
\end{align}
Projecting the tensor equations with $q^A q^B$, the vector equations with $q^A$, and writing the contractions in scalar equations using Eq.~(\ref{metric-to-dyad}), we re-express the equations above in terms of spin-weighted scalars. Note also that we incorporate the $\eth$-formalism by defining $\eth = q^A \nabla_A$ and $\bar{\eth} = \bar{q}^A \nabla_A$ \cite{Goldberg:1966uu, Gomez_1997}.

We again group the equations in different orders of the coupling parameter $\epsilon$:

\begin{itemize}
\item Order $\mathcal{O}(\epsilon)$:
\begin{align}
    \partial_v c &= 2M \partial_v s + \frac{1}{2} s  - \frac{1}{2M} c - 2 M \eth h \\[5pt]
    \partial_v h &=  \eth k - \frac{1}{8M^2} \bar{\eth} \left(\partial_v c\right) \\[5pt]
    k &= -\frac{1}{8M} \left(\eth \bar{h}+\bar{\eth}h \right)+\frac{1}{64 M^3}\left(\eth^2 \bar{c}+\bar{\eth}^2c\right)
\end{align}
\begin{widetext}
\item Order $\mathcal{O}(\epsilon^2)$:
\begin{align}
    \partial_v \vartheta &= \frac{1}{4M} \vartheta - \frac{1}{32M^4} \left( \partial_v c \right) \left(\partial_v \bar{c}\right) - \left( \partial_v \varphi \right)^2 \\[5pt]
    \partial_v C &= 2M \partial_v S + \frac{1}{2} S  - \frac{1}{2M} C + 2k \left[M s - c \right]  + 4M \left(\eth H  + h^2 \right) + 2M \left( \eth \varphi \right)^2 \label{eq:cequation}\\[5pt]
    \partial_v H &=  \eth K- \frac{1}{16M^4} \left[ 2M^2\bar{\eth}\left(\partial_v C \right)  
    + {\eth} \left(c \, \partial_v \bar{c}\right)\right]   + \frac{1}{2} \eth \vartheta + \partial_v \varphi \eth \varphi \label{eq:hequation} \\[5pt]
    K &= -\frac{1}{32M^4} \partial_v \left( c \bar{c}\right) + \frac{1}{32M^3} \left[ s  \bar{s} + \left(\partial_v\bar{c}\right)  s +c \, \partial_v\bar{s} + \bar{c} \, \partial_v s\right]  \notag \\ & \quad - \frac{1}{128M^5} \left[2 \bar{c}c + M(s \bar{c} + c \bar{s} ) \right]
     - \frac{1}{8M} \left[ \eth \bar{H} + \bar{\eth} H + 2 h \bar{h}\right] - \frac{1}{4M} \eth \varphi \bar{\eth} \varphi \notag \\
     & \quad + \frac{1}{64M^3} \left[\bar{\eth}^2 C + \eth^2 \bar{C} + \frac{1}{8} \left(\eth c \bar{\eth} \bar{c} - \eth \bar{c} \bar{\eth} c + 8 c\bar{c} \right) \right]
\end{align}
\end{widetext}
\end{itemize}
Finally, we express all the terms above in terms of spin-weighted spherical harmonics \cite{10.1063/1.1931221, Goldberg:1966uu}.
As an example, we decompose the scalar field as follows:
\begin{align}
    \Phi^{(1)} \equiv \varphi = \sum_{\ell = 0}^N \sum_{m=-\ell}^{\ell} {\varphi_{\ell m}\left(v\right)} \, {}_0Y_{\ell m} \,,
\end{align}
where $\varphi_{\ell m}\left(v\right)$ is the spin-weighted coefficient indexed by integers $\{\ell, m \}$ and ${}_0Y_{\ell m}$ denotes the spherical harmonic with spin weight 0. Similarly, vectors and tensors are decomposed into spin-weighted coefficients and spherical harmonics of spin weights 1 and 2, respectively.

We solve the resulting equations given in Appendix~\ref{appendix:numerical} for the spin coefficients using numerical methods.

\subsection{Numerical Results \label{secV:level3}}

We solve the horizon equations recast in the spin-weighted formalism with a fourth-order Runge--Kutta integrator in Python.\footnote{We make extensive use of the Python libraries NumPy~\cite{harris2020array}, Matplotlib~\cite{Hunter:2007}, SymPy~\cite{10.7717/peerj-cs.103}, and SciPy~\cite{2020SciPy-NMeth}.} We obtain the time evolution of the spin-weighted scalars $\{\vartheta, H, C,K\}$, from which we recover the respective geometric quantities on the horizon, $\{ \Theta, \mathcal{H}_A, q_{AB}, \kappa\}$. 

The system of equations at hand consists of four coupled ordinary differential equations. Due to the teleological nature of the event horizon, the system is set up as a Schwarzschild black hole at a large null time $v$, and the boundary condition that the perturbations diminish at large $v$ is imposed. Then, the equations are solved from the future to the past of the evolution. (See \cite{G_mez_2001} for more details of ``backward-in-time" evolution.) Therefore, the boundary conditions imposed at the distant future of the horizon are that the null expansion and the Hájíček field vanish at all perturbative orders. At the same time, the 2-metric becomes the metric on the unit sphere, and the surface gravity assumes its value for a static Schwarzschild black hole. 

\begin{figure}
    \centering
    \includegraphics[width=\linewidth]{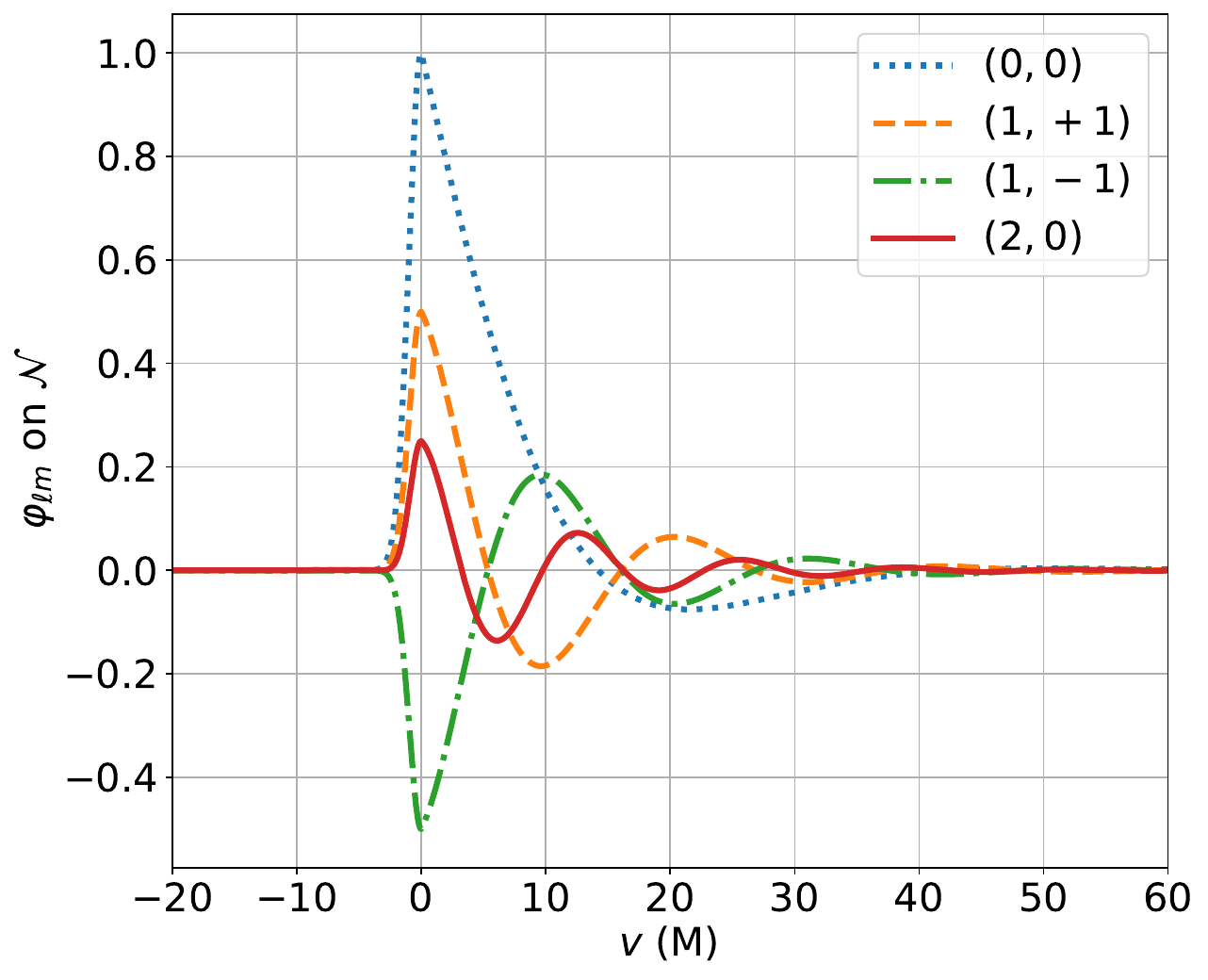}
    \caption{Time profile of different $(\ell,m)$ modes on the horizon. Peak amplitudes, $\mathcal{A}_{\ell m}$, are set to satisfy: 1 = $\mathcal{A}_{00} = 2 \mathcal{A}_{11}  = - 2 \mathcal{A}_{1-1} = 4 \mathcal{A}_{20}$. The negative amplitude is needed for the $\left( \ell,m \right) =(1,-1)$ to ensure the scalar field is real. See Appendix~\ref{appendix:scalarfield} for details on the construction of this profile.}
    \label{fig:source}
\end{figure}

The source is modeled as a massless scalar field, the time profile of which \textit{on the horizon} overlays a Gaussian pulse and a ringdown phase with quasinormal mode (QNM) frequencies of a Schwarzschild black hole under scalar perturbations. 
\begin{align}
    {\varphi_{\ell m}\left(v\right)} \propto \exp{\left(-(v-v_0)^2/(2\sigma)\right)} \exp{\left( -i\omega_{\ell m} v\right)} 
\end{align}
where $v_0$ is the location of the peak, $\sigma$ is its width, and $\omega_{\ell m}$ are the complex QNM frequencies.
In our toy model, we choose the amplitude of the source to peak at $v_0 = 0$ and set its standard deviation to $\sigma=1$. We restrict the initial data to the spherical harmonic modes $\left(\ell,m\right)\in\left\{(0,0), (1,-1),(1,1),(2,0)\right\}$. The time profile of the scalar field \textit{on the horizon} is shown in Fig.~\ref{fig:source}. Details of this implementation are left to the Appendix~\ref{appendix:scalarfield}. Note that the $m\neq0$ modes allow breaking of axisymmetry in the perturbed system from which we recover a notion of angular momentum as discussed below.

\begin{figure*}[htb]
  \hspace{-2cm}
  \includegraphics[width=\textwidth]{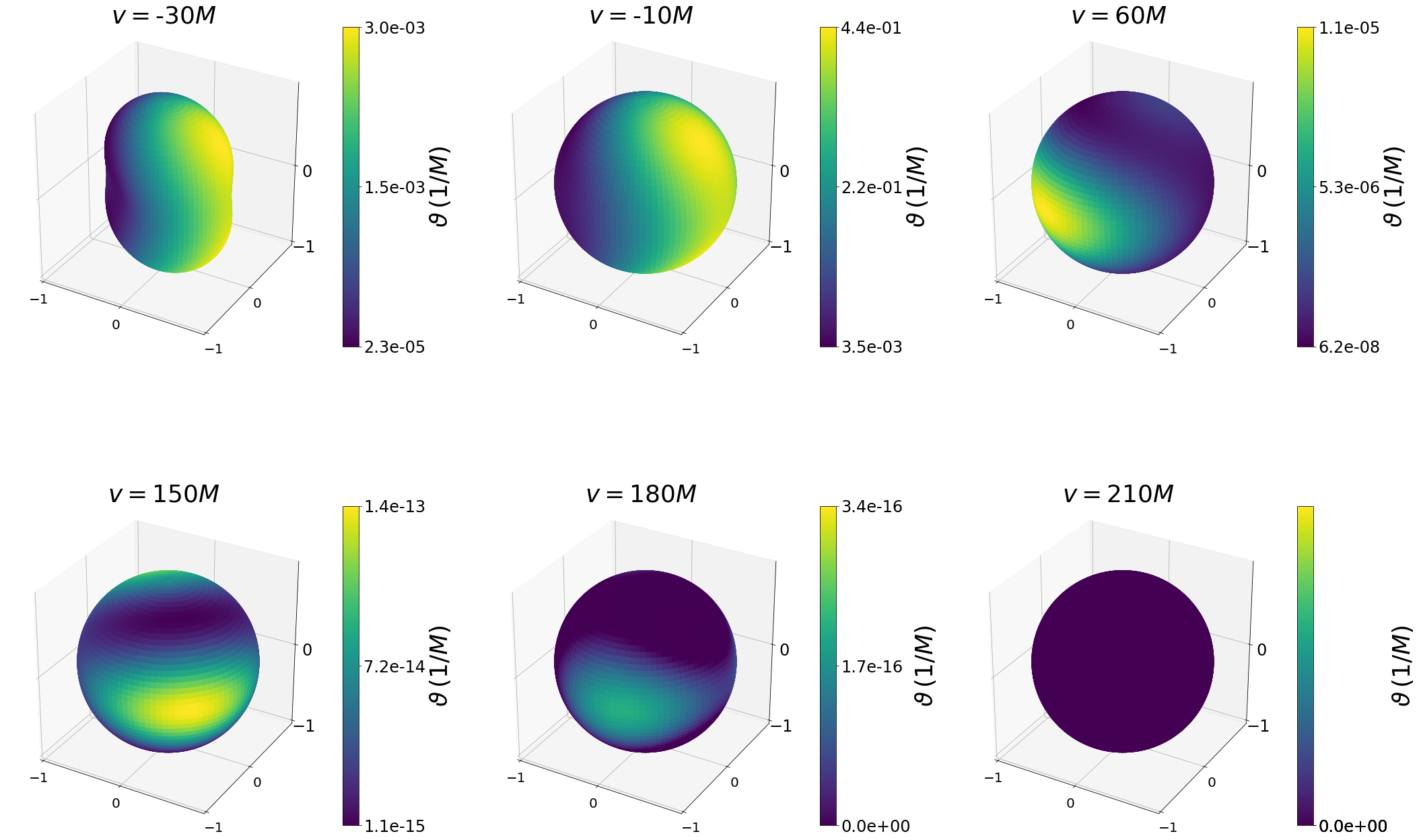}
  \caption{Horizon 2-sphere embedding at different times. Radial coordinate is scaled with $\sqrt{\det{q_{AB}}}$ to illustrate the deformations of the surface. The peak of the scalar field is reached at $v = 0$. The color map refers to the value of the null expansion in $\mathcal{O}\left(\epsilon^2\right)$, $\vartheta\left(\theta,\phi\right)$, evaluated on a grid of angles on the 2-sphere. A cut-off value of $10^{-16}$ is used in the color map. That is, $\vartheta\left(\theta,\phi\right)$ is set to 0 for any cell on the grid on which the value of null expansion is less than $10^{-16}$. The perturbative parameter is set to satisfy   $\epsilon^2 = 3 \times 10^{-8}$, which is chosen such that the deformations on the metric are visible in the embedding plot at early times.}
  \label{fig:snapshots}
\end{figure*}

In Fig.~\ref{fig:snapshots}, we illustrate embeddings of the horizon 2-sphere in three dimensions at different times during the horizon's evolution.
As explained above, the simulation is initiated at the future end time, which is set to the value $v_f = 210 \mathrm{M}$ (final time) in units of the final mass of the black hole; then, the horizon equations are integrated backward in time. The horizon becomes more deformed in the past of the scalar field peak, as it anticipates the infalling matter. The total duration of the simulations is chosen for convenience to demonstrate the deformations at small $v$ and relaxation to the Schwarzschild solution at large $v$. This evolution is within the perturbative regime as the total mass change is less than $0.1\%$ of the final black hole mass.

\subsubsection{Quadratic quasinormal modes \label{secV:level3.1}}

Quadratic quasinormal modes have been of interest for the study of nonlinear effects in gravity \cite{Ma_2024, Bucciotti_2024, Khera_2025, Khera_2023}. Here, we focus on the generation of quadratic modes from the linear modes of source fields on the black hole event horizon.

\begin{figure}[htb]
\hspace*{-1cm}
    \includegraphics[width=0.5\textwidth]{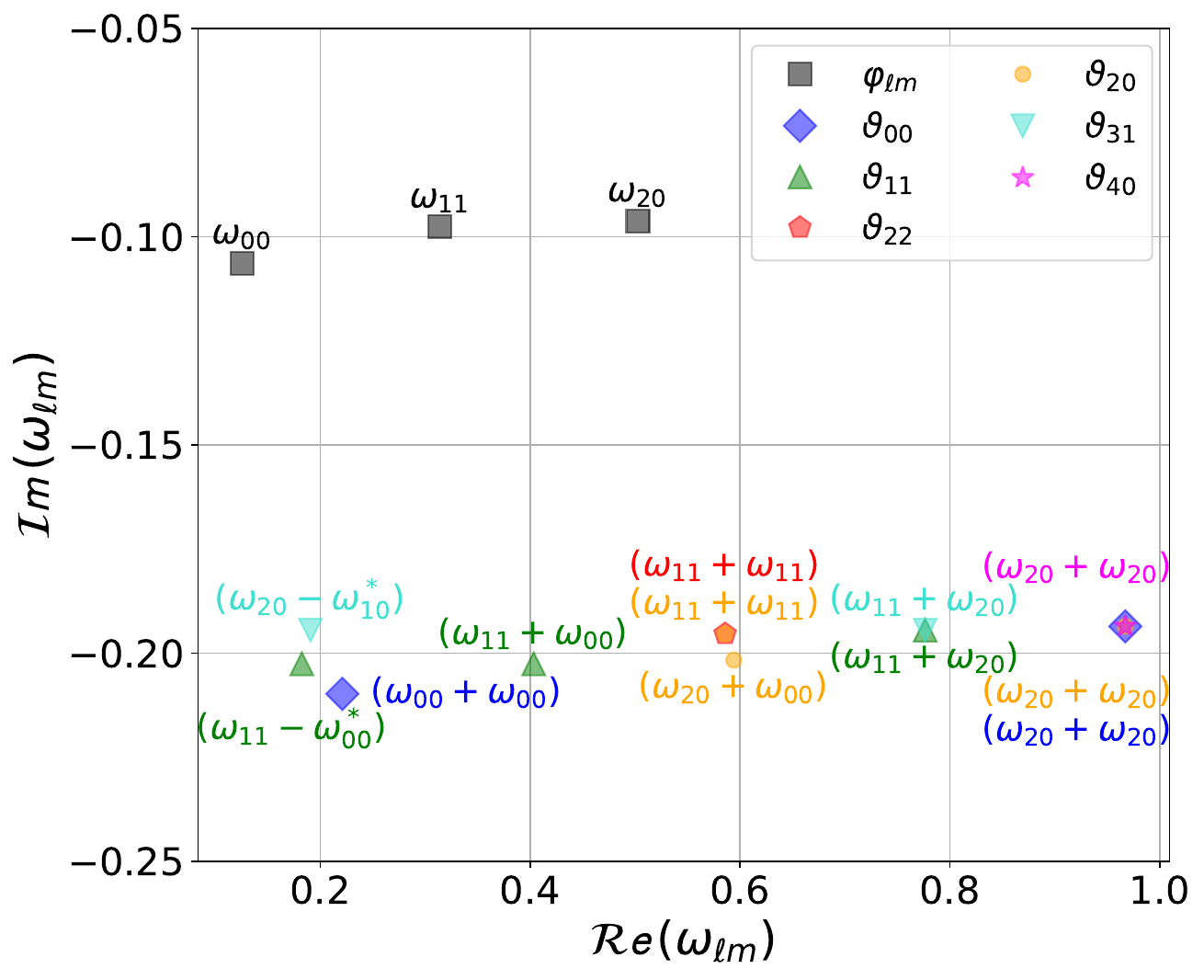}
    \caption{Real and imaginary parts of the quasinormal mode (QNM) frequencies extracted from the evolution of the null expansion in $\mathcal{O}\left(\epsilon^2\right)$. Grey squares denote the input QNM frequencies of the scalar field source. Note that QNM frequencies of $\vartheta_{\ell m}$ that appear multiple times are generated by different parent modes. Parent modes are written near the markers in the same color as the markers.}
    \label{fig:qnm}
\end{figure}

QNM frequencies of the scalar field source modes combine quadratically through the horizon equations. This follows from the fact that the stress-energy tensor is quadratic in the scalar field, so are the corresponding source terms, $ S_i \propto (\partial_v\Phi \partial_i\Phi)$. Therefore, we expect these quadratic QNM frequencies to appear in the evolution of the geometric quantities. We extract the frequencies from our result using the QNM rational filter \cite{Ma_2022, Ma_2023, Ma_2023_2}. These results are summarized in Fig.~\ref{fig:qnm}. 

Notice that a quadratic QNM can have different couples of parent modes. For example, as shown in Fig.~\ref{fig:qnm}, the expansion mode $\vartheta_{11}$ can be generated by the combinations of sources involving $\varphi_{11}\varphi_{00}$ and $\varphi_{11}\varphi_{20}$. Also, notice that the same parent modes can be combined in different ways to generate the daughter mode. Observe the combinations of $\omega_{11}+\omega_{00}$ and $\omega_{11}-\omega_{00}^*$ in $\vartheta_{31}$. Here, ${}^*$ denotes the complex conjugate. 

To further confirm that these modes are indeed quadratic QNMs, we check the amplitudes as we change the amplitude of the scalar field profile. For example, we observe that the amplitude of $\vartheta_{40}$ scales quadratically with the amplitudes of $\varphi_{20}$, further supporting that these modes are generated as quadratic combinations of linear modes. 

\subsubsection{Horizon area \label{secV:level3.2}}
We quantify the strength of the scalar field perturbations by tracking the change in the horizon area, which is also related to the mass of the black hole.

The surface area of the horizon can be calculated as
\begin{align}
    A  &= \int_\mathcal{S} \sqrt{q} \, \mathrm{d}\theta \, \mathrm{d}\phi \, .
\end{align}
where $q \equiv \det q_{AB}$ and the integral is evaluated over the (deformed) 2-sphere $\mathcal{S}$.
Taking a derivative with respect to the null time, the surface area rate of change can be related to the null expansion.
\begin{align}
    \frac{\mathrm{d}A}{\mathrm{d}v} &=\int_\mathcal{S}\frac{\partial_v  q }{2q} \sqrt{q}  \, \mathrm{d}\theta \, \mathrm{d}\phi \, \, \\
    &= \int_\mathcal{S} \Theta \sqrt{q} \, \mathrm{d}\theta \, \mathrm{d}\phi \, .
\end{align}
Here, we used 
\begin{align}
    \Theta \equiv \frac{1}{2}q^{AB} \partial_v q_{AB} = \frac{\partial_v q}{2 q} \, . 
\end{align}
In the perturbative solution, this becomes
\begin{align}
    \frac{\mathrm{d}A}{\mathrm{d}v} = \epsilon^2  (2M)^2 \int_\mathcal{S}  \vartheta  \sin{\theta} \, \mathrm{d}\theta \, \mathrm{d}\phi \, .
\end{align}

On the other hand, the energy flux due to the scalar field passing through the horizon is given by \cite{Poisson_2004}
\begin{align}
    \frac{\mathrm{d}E}{\mathrm{d}v} &= \int_S \mathcal{T}_{ab} l^a l^b \sqrt{q} \, \mathrm{d}\theta \, \mathrm{d}\phi \\
    &=\int_S \mathcal{T}_{vv} \sqrt{q} \, \mathrm{d}\theta \, \mathrm{d}\phi  \, ,
\end{align}
which, in the perturbative scheme, becomes
\begin{align}
 \frac{\mathrm{d}E}{\mathrm{d}v}= \epsilon^2  (2M)^2\int_S \left( \partial_v\varphi \right)^2 \sin{\theta} \, \mathrm{d}\theta \, \mathrm{d}\phi \, .
\end{align}
where $M$ is the final black hole mass.

Since the perturbation parameter $\epsilon$ is small, the change in the black hole parameters through the entire time evolution can be assumed quasistatic.
In this regime, one can relate these two expressions via the first law of black hole mechanics\footnote{See~\cite{Wald_2001} for a review on black hole thermodynamics.} 
\begin{align}
    \mathrm{d} E = \frac{\kappa}{8\pi} \mathrm{d}A + \Omega_H \mathrm{d}J \, .
\end{align}

In the perturbed black hole model we study here, the unperturbed solution is a Schwarzschild black hole. Therefore, the second term on the right-hand side, which denotes the angular velocity and momentum of the black hole horizon, is of $\mathcal{O}\left(\epsilon^4\right)$. Up to the second order, we can express
\begin{align}
    \frac{\mathrm{d} E}{\mathrm{d}  v} &= \frac{1}{32 \pi M } \frac{\mathrm{d} A}{\mathrm{d}  v}  + \mathcal{O}\left(\epsilon^4\right) \, , 
\end{align}
where we used $\kappa^{(0)}=1/\left(4M\right)$.

\begin{figure}[htb]
    \centering
    \includegraphics[width=0.5\textwidth]{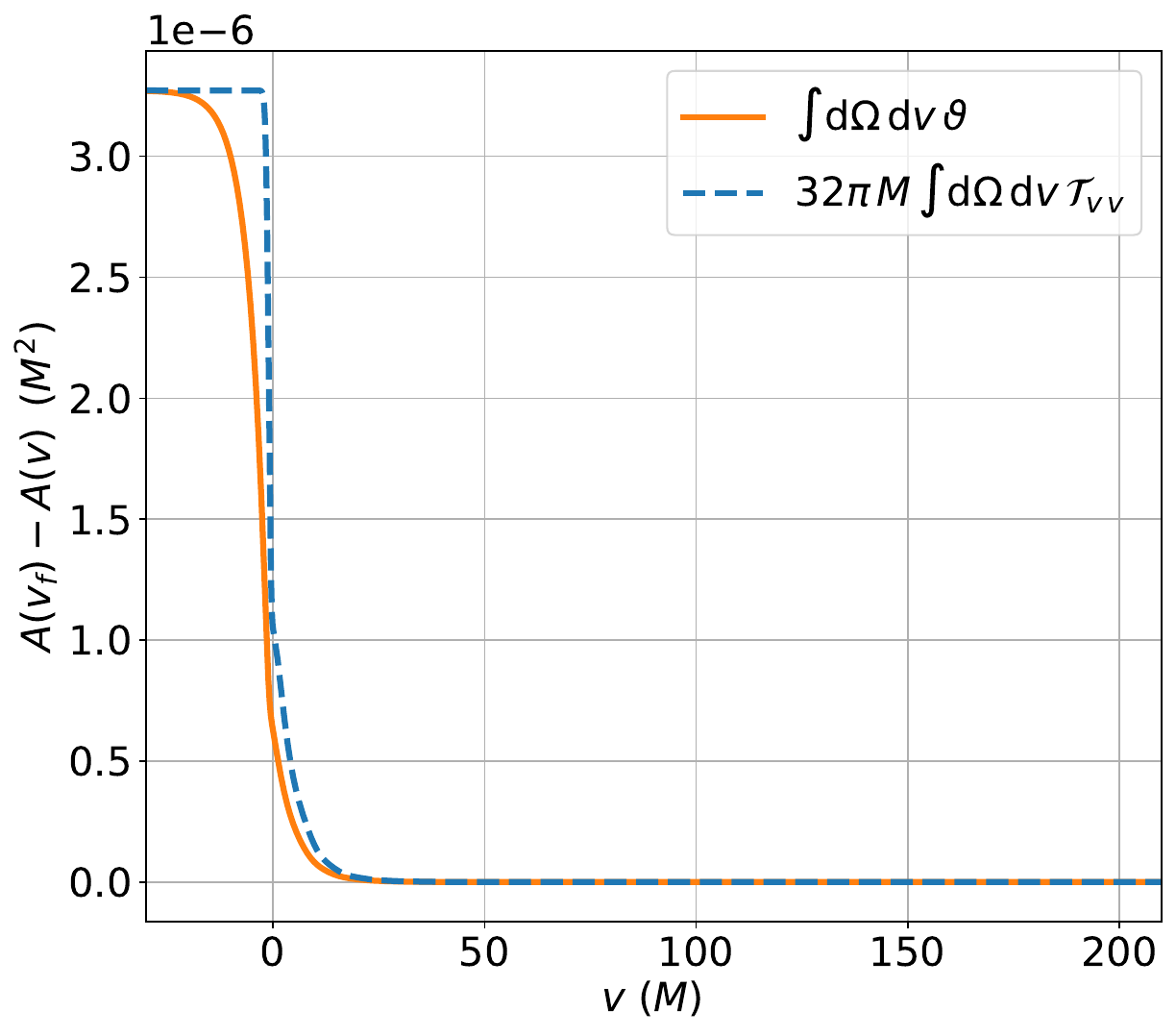}
    \caption{Difference between the final black hole area at time $v_f$ and the area at time $v$, computed from the null expansion (solid orange curve) and the energy flux (dashed blue curve) versus time. The perturbative parameter is the same as in Fig.~\ref{fig:snapshots}, which satisfies $\epsilon^2 = 3 \times 10^{-8}$.}
    \label{fig:area}
\end{figure}

In Fig.~\ref{fig:area}, we plot the change in the horizon area with respect to the final area at $v = v_{f}$ alongside the flux of energy carried by the scalar field. The total change in the area from the beginning of the perturbations to the end is $3.5 \times 10^{-6} M^2$ where $M$ denotes the final mass.

The two curves in Fig.~\ref{fig:area}, closely follow each other at later times. Before the peak of the perturbation, at $v=0$, the change in the area calculated from the scalar field stress-energy tensor (dashed blue curve) stays constant, and it starts to change abruptly as the perturbation gets stronger. On the contrary, the area change computed from the null expansion (solid orange curve) varies smoothly before $v=0$. This behavior of the null expansion is due to the teleological nature of the event horizon.

\subsubsection{Horizon angular momentum \label{secV:level3.3}}
Another physical quantity that can be extracted from the simulations is the angular momentum carried by the scalar field. As the black hole relaxes to a Schwarzschild solution at late times, after the perturbations die off, one can think of the non-axisymmetric modes in the scalar field source as carrying angular momentum of the opposite sign to the initially spinning black hole. 

The flux of angular momentum through the event horizon is calculated as follows \cite{Poisson_2004}:
\begin{align}
    \frac{\mathrm{d}J}{\mathrm{d}v} &= -\int_S \mathcal{T}_{ab} l^a \psi^b \sqrt{q} \, \mathrm{d}\theta \, \mathrm{d}\phi \, , \\
    &=-\int_S \mathcal{T}_{v\phi} \sqrt{q} \, \mathrm{d}\theta \, \mathrm{d}\phi \, .
\end{align}
Assuming that the deviation from axisymmetry is small through the perturbative evolution, the axial Killing vector in the perturbed spacetime is expressed by $\psi^a \equiv \partial_\phi$. Note that these expressions hold within the assumption that the perturbed black hole deviates slightly from the Kerr solution as discussed in~\cite{Poisson_2004}. In the perturbative expansion, the above expression becomes
\begin{align}
     \frac{\mathrm{d}J}{\mathrm{d}v}  &= -\epsilon^2  (2M)^2\int_S \left( \partial_v\varphi \partial_\phi\varphi  \right) \sin{\theta} \mathrm{d}\theta \mathrm{d}\phi \, .
\end{align}

On the other hand, the DNS equation can be used to relate the Hájíček field to the angular momentum on the event horizon through Damour's original prescription \cite{Damour1982}
\begin{align} \label{eq:angmomhaji}
    J &= -\frac{1}{8\pi}\int_S \mathcal{H}_{A} \psi^A \mathrm{d}\Omega  \, .
\end{align}

\begin{figure}[htb]
    \centering
    \includegraphics[width=0.5\textwidth]{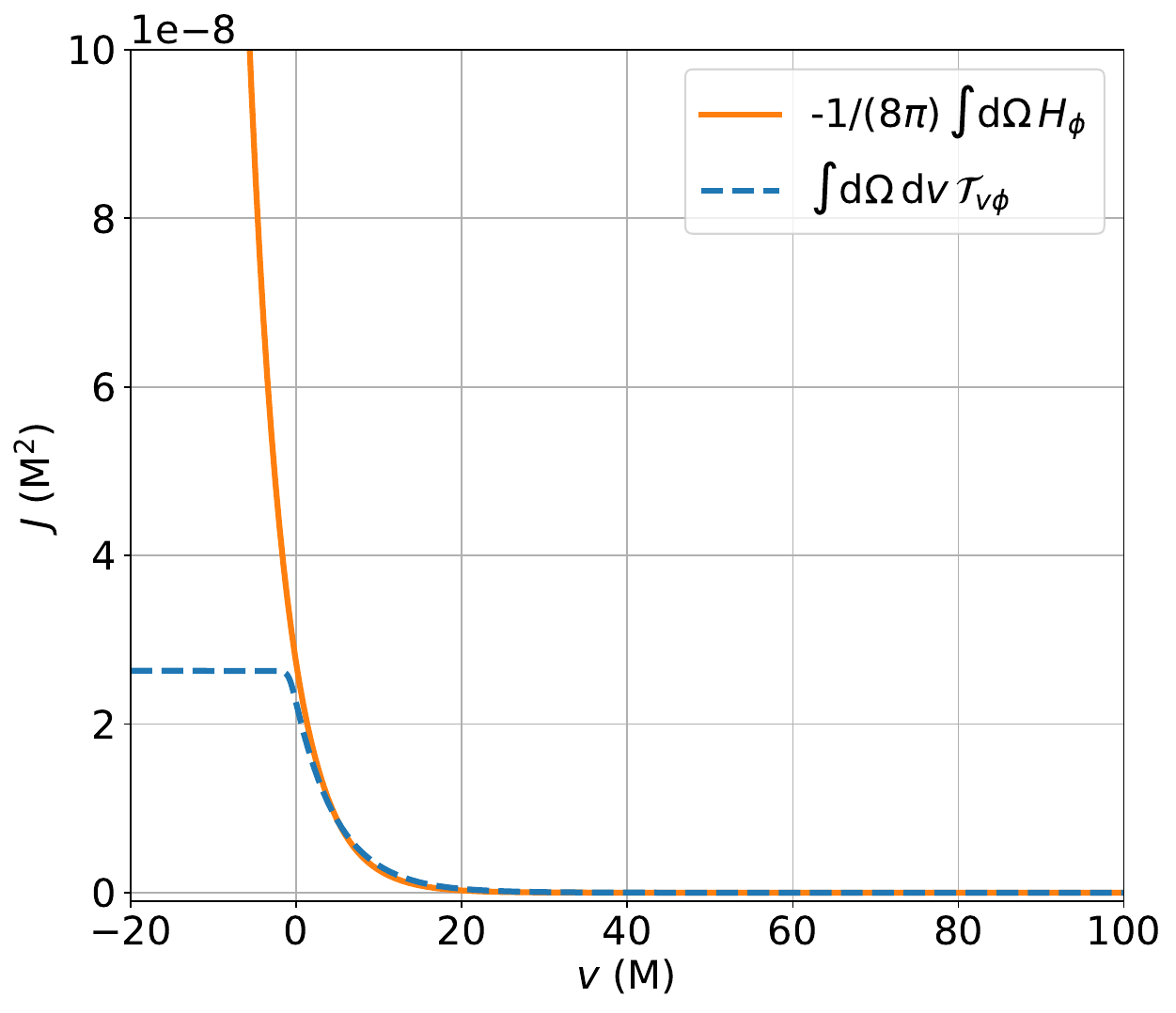}
    \caption{Angular momentum of the horizon computed by integrating the Hájíček field (solid orange curve) and integrating the angular momentum flux due to the scalar field (dashed blue curve). The perturbative parameter is the same as in Fig.~\ref{fig:snapshots}, which satisfies $\epsilon^2 = 3 \times 10^{-8}$.}
    \label{fig:angmom}
\end{figure}

In Fig.~\ref{fig:angmom} we plot the horizon angular momentum computed from the Hájíček field and the scalar field. As was the case for the horizon area plotted in Fig.~\ref{fig:area}, the two curves denoting the angular momentum of the black hole in Fig.~\ref{fig:angmom} agree at late times. Before the peak of the perturbation at $v=0$, the dashed blue curve computed from the flux due to the scalar field stress-energy tensor implies that the horizon has a constant angular momentum. Then, the scalar field deposits angular momentum of the opposite sign to the black hole, such that the black hole becomes non-spinning at the end of its evolution. 

Contrarily, the angular momentum computed from the Hájíček field has a completely different behavior. The solid orange curve rises exponentially before $v=0$. This is due to the secular terms in Eq.s~\ref{eq:cequation}~\&~\ref{eq:hequation}, which permeate to the angular momentum computed from Eq.~\ref{eq:angmomhaji}. Therefore, at some $v$ around $v=0$, the perturbative approximation fails to capture the physical behavior.

Note that our analysis below on the equilibration relies only on the evolution of the null expansion, which agrees with the expected change in the area as shown above in Fig.~\ref{fig:area}.

\subsubsection{Equilibration on the horizon \label{secV:level3.4}}
\begin{figure*}[htb]
  \includegraphics[width=\textwidth]{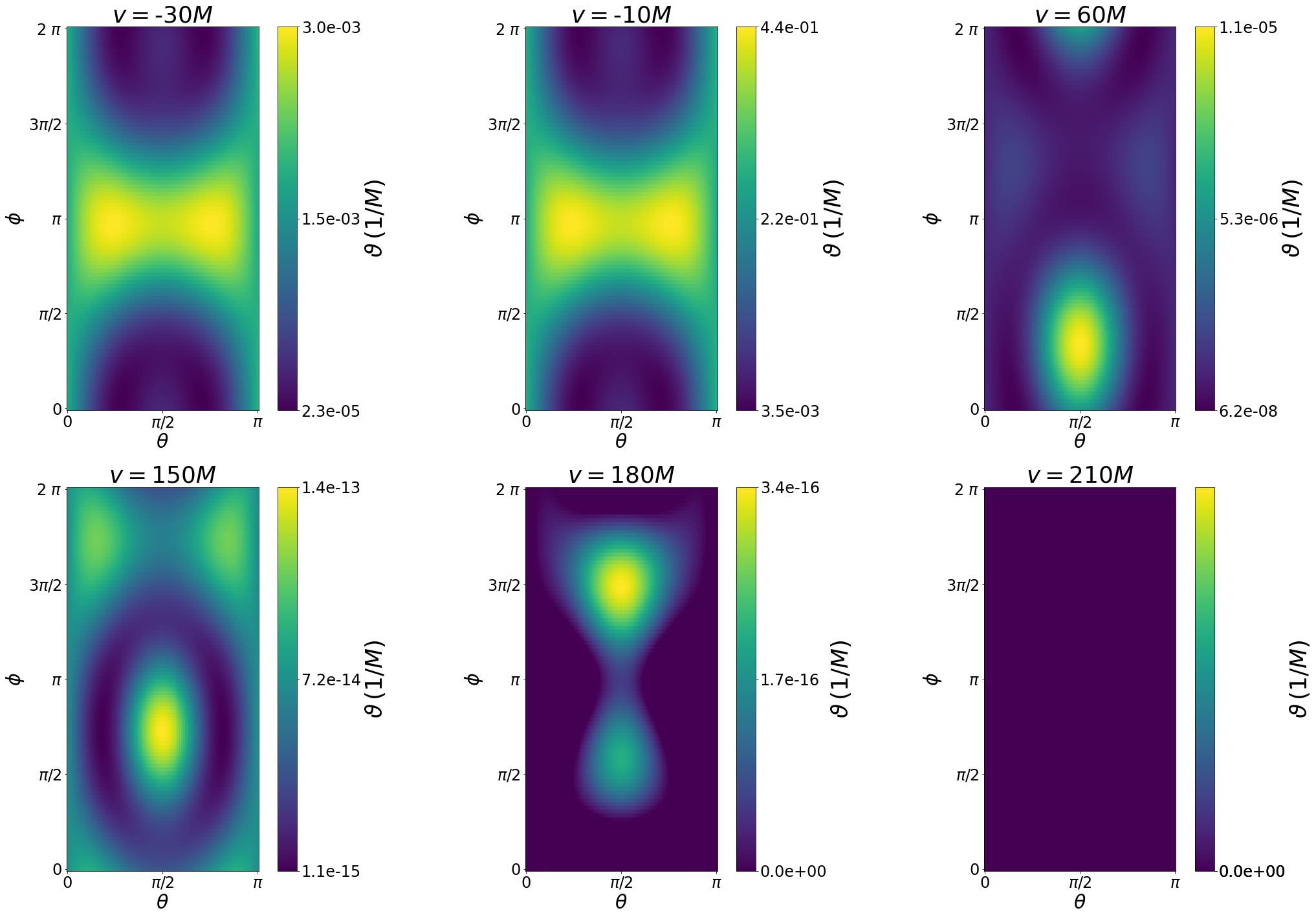}
  \caption{Snapshots of null expansion of the horizon in $\mathcal{O}\left(\epsilon^2\right)$ plotted with respect to the angles on the sphere, $\vartheta\left(\theta,\phi\right)$ This quantity is related to the energy density $\mathcal{E}$ of the dual Carrollian fluid. A cutoff value of $10^{-16} M^{-1}$ is applied, below which $\vartheta$ is set to 0. The perturbative parameter is the same as in Fig.~\ref{fig:snapshots}, which satisfies $\epsilon^2 = 3 \times 10^{-8}$.}
  \label{fig:temperature}
\end{figure*}

In Fig.~\ref{fig:temperature}, we plot the time evolution of the null expansion $\Theta{\left(v,\theta,\phi\right)}$ at different polar angles on the sphere. We observe that the expansion is nonzero even at early times due to the teleological behavior of the event horizon. Later, it decays following the ringdown of the scalar field driving its evolution. We employ a cutoff value of $10^{-16}  M^{-1}$ (where $M$ is the final mass of the black hole), below which we set the expansion to 0. 

Through the identification of the horizon dynamics with the Carrollian fluid equations, the behavior of the null expansion on the horizon is tied to the energy density of the Carrollian fluid dual to it. Thus, we interpret Fig.~\ref{fig:temperature} as the equilibration of the dual Carrollian fluid by the expansion of the geometry on which the fluid lives. That is, as the perturbed black hole horizon expands and approaches a static Schwarzschild solution, its Carrollian fluid dual loses energy and equilibrates. Note that the endpoint of the equilibration is reached once the horizon becomes static; the Carrollian dual reaches zero energy density.

\subsubsection{Vorticity of  $\mathcal{H}_A$ \label{secV:level3.5}}

Finally, we compute the vorticity of the Hájíček field defined by \cite{Redondo_Yuste_2023}
\begin{align}
    \omega^{\mathcal{H}}_{AB} \equiv \nabla_{[A} \mathcal{H}_{B]}.
\end{align}
Hájíček vorticity is informative in the fluid description as the Hájíček field is identified with the fluid momentum through the Carrollian fluid/horizon dictionary. 

\begin{figure*}[htb]
  \centering
  \includegraphics[width=\textwidth]{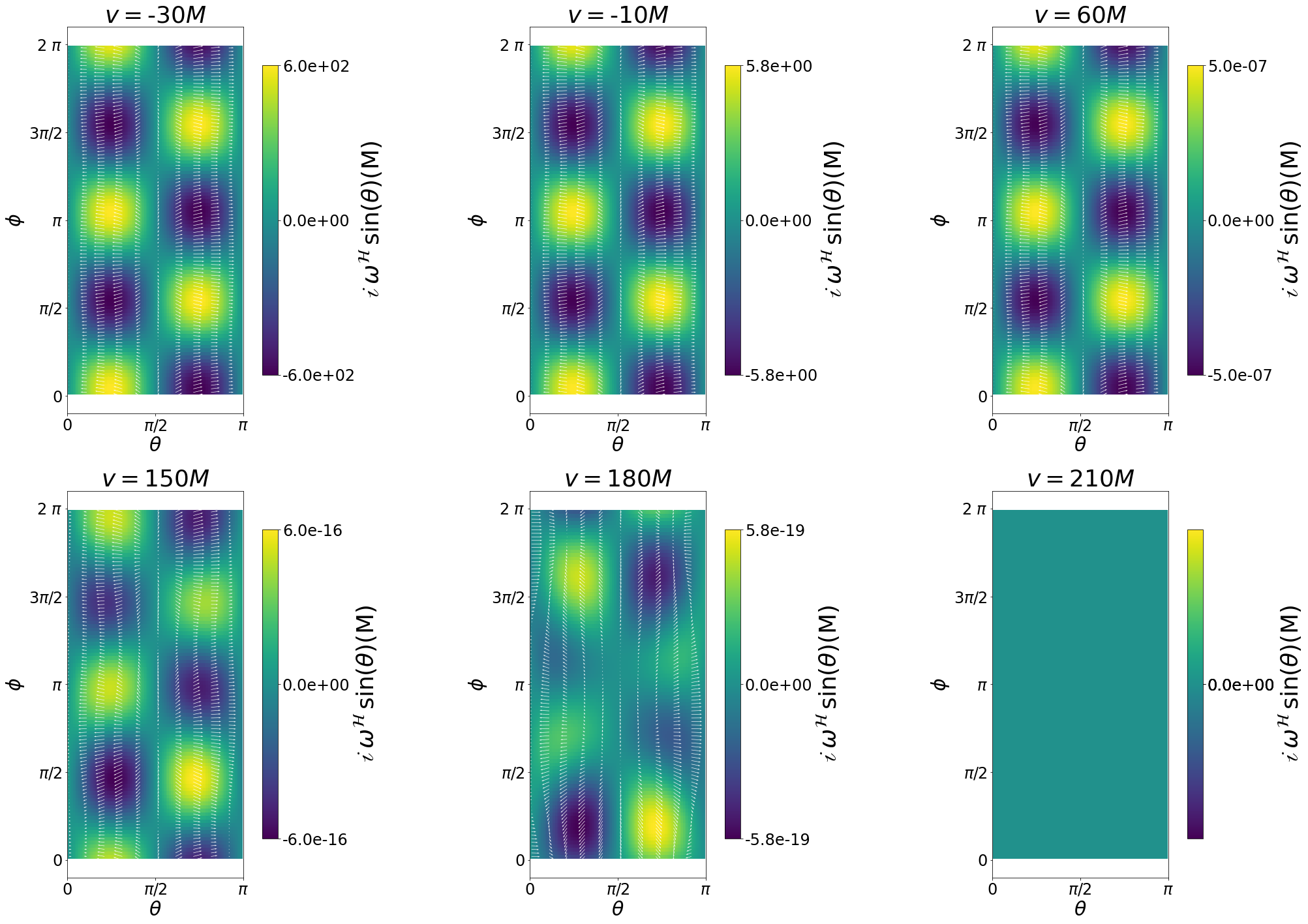}
  \caption{\textbf{Background color plot:} Snapshots of angular projection of Hájíček vorticity, multiplied by $\sin\theta$, $i\omega^{\mathcal{H}}\left(\theta, \phi\right) \sin\theta$, to correct for the distortions near the poles $\phi = 0, \, 2\pi$.  plotted with respect to the angles on the sphere. Color values are normalized in each snapshot. \textbf{ Foreground quiver plot:} Components of the Hájíček vector, $\mathcal{H}^A$, along the axial and azimuthal angles. Arrow lengths are normalized in each snapshot. For both plots, the expansion parameter is set to satisfy $\epsilon^2 = 3 \times 10^{-8}$.}
  \label{fig:vortices}
\end{figure*}

In Fig.~\ref{fig:vortices}, we plot the $\mathcal{O}\left(\epsilon^2\right)$ contributions to the angular projection of the vorticity $\omega^{\mathcal{H}} \equiv  q^A q^B \omega^{\mathcal{H}}_{AB} / 2$ and the vector components of $\mathcal{H}^A$ along the directions on the 2-sphere, $\{\mathcal{H}^\theta, \mathcal{H}^\phi \}$. The background color map in Fig.~\ref{fig:vortices} highlights the angular structure of Hájíček vorticity. On the other hand, the quivers represent the directions of $\mathcal{H}^A$ that trace the deformations on the surface. The axial component also carries information about the angular momentum of the horizon, as discussed above. We leave detailed discussions on the Hájíček field, horizon momentum, and the vorticity of the dual Carrollian fluid to future work.

\section{\label{secVI:level1} Conclusions}
In this paper, we studied the dynamical properties of Carrollian fluids dual to a perturbed black hole event horizon. Within the program of Carrollian fluid/horizon duality, we worked out the dynamics of a black hole event horizon perturbed by matter fields in the bulk to investigate the behavior of the Carrollian fluid under sources. On the gravity side, equations describing the horizon dynamics coupled to a matter field are analytically derived. These horizon equations are then numerically solved for a toy model of a Schwarzschild black hole perturbed by a non-axisymmetric distribution of a massless real scalar field near its horizon. On the Carrollian fluid side, we interpret the relaxation of the horizon after the perturbation is turned off as a mechanism for the dual Carrollian fluid to equilibrate. 

 Fluid ``particles" cannot move in space in the Carrollian limit. However, the analogous dynamics of the horizon demonstrate that the null expansion vanishes at  $v \rightarrow \infty$. The Carrollian fluid/horizon dictionary identifies the fluid energy density with the null expansion of the horizon.
 Therefore, the Carrollian fluid analogous to the event horizon equilibrates via modifying the geometry on which it lives.
 
\textcolor{black}{The full equations describing horizon dynamics are nonlinear. Therefore, one expects that the Carrollian fluid/horizon geometry duality should extend to the general case beyond a perturbative treatment. Nevertheless, we observe the hints of nonlinear effects, such as the generation of quadratic QNMs on the horizon, which can provide us with intuition for the nonlinear effects within fluid/gravity correspondence. Resolving the implications of this duality in a full dynamical setting would require a more sophisticated numerical approach and is left for future work.} 
This could include a dynamical approach to \textit{hydrodynamical horizons} \cite{freidel2024geometrycarrollianstretchedhorizons}, which is briefly discussed at the end of Appendix~\ref{appendix:bgauge}. Another natural extension of this work is to build on the discussions in \cite{Redondo_Yuste_2023} to study the nonlinear-to-linear transition and explore enstrophy definitions within Carrollian fluids and the counterpart in gravity, if such a concept exists for such fluids. 

A formalism similar to this work can be used to investigate the Carrollian fluid dual to any other null surface at null infinity \cite{Ciambelli:2025mex} or null surfaces in the bulk not corresponding to an event horizon.  Notice that, in such a case, the surface will always expand (if outside the horizon) or always contract (if inside of it), and thereby the associated Carrollian energy will not equilibrate.
Finally, on the black hole perturbation front, the horizon equations we studied in this work can yield insights about the dynamics of the horizon
and its relation to quasinormal mode behavior and the nonlinear-to-linear transition (see, for example, \cite{May_2024}). \

\begin{acknowledgments}
We thank Luca Ciambelli, Puttarak Jai-akson, Jaime Redondo-Yuste, and Sizheng Ma for useful discussions.
   LL acknowledges support from the Natural
Sciences and Engineering Research Council of Canada
through a Discovery Grant. LL also thanks financial support via the Carlo Fidani Rainer Weiss Chair at Perimeter
Institute and CIFAR. This research was supported in part
by Perimeter Institute for Theoretical Physics. Research
at Perimeter Institute is supported in part by the Government of Canada through the Department of Innovation,
Science and Economic Development and by the Province
of Ontario through the Ministry of Colleges and Universities. 
\end{acknowledgments}

\section*{Data availability}
Our code is publicly available \cite{CodeOnGit}. 

\appendix
\section*{Appendices}
\textcolor{black}{
Some details of the computations and the arguments of the main text are included in the appendices. In Appendix~\ref{appendix:derivatives}, we summarize the derivatives and connections relevant to the stretched Carrollian structure.  In Appendix~\ref{appendix:gaugechoices}, we expand on the gauge choices implemented in Section~\ref{secII:level1}. We leave some expressions in the derivations of perturbative equations to Appendix~\ref{appendix:equationsO2}, while expounding the numerical implementation in Appendix~\ref{appendix:numerical}. Finally, in Appendix~\ref{appendix:examplegauge}, we explicitly treat two well-known spacetimes in the structure introduced in Section~\ref{secII:level1} as concrete examples.}

\section{Derivatives \& connections \label{appendix:derivatives}}

The prescription \cite{freidel2024geometrycarrollianstretchedhorizons} we followed to construct a null hypersurface involved projecting quantities in the bulk onto different hypersurfaces, as discussed in the main text. Here, we summarize how the covariant derivatives and curvature connections are induced in such hypersurfaces.

The rigged derivative $D_i$ is defined  \cite{freidel2022carrollian} as the projection of the covariant derivative $\nabla_a$ on $\mathscr{M}$ onto $\mathscr{H}$. For a tensor with mixed indices on $\mathscr{H}$, the rigged derivative acts as the following
\begin{equation}
    D_i X_j^{\hspace{4pt}k} \equiv \Pi_i^{\hspace{4pt}d} \Pi_j^{\hspace{4pt}e} \left( \nabla_d X_e^{\hspace{4pt}f} \right) \Pi_f^{\hspace{4pt}k}
\end{equation}
A different notion of derivative is defined for tensors on $\mathcal{S}$ through
\begin{equation}
    \hat{\nabla}_A X_B^{\hspace{4pt}C} \equiv q_A^{\hspace{4pt}i} q_B^{\hspace{4pt}j} \left( D_i X_j^{\hspace{4pt}k} \right) q_k^{\hspace{4pt}C}
\end{equation}
where it is useful to note $ q_A^{\hspace{4pt}a} = \left( e_A \right)^a$. This derivative satisfies $\hat{\nabla}_C \, q_{AB} = 0$ and preserves Carrollian diffeomorphism; therefore, it is often referred to as the Levi-Civita--Carroll covariant derivative \cite{Ciambelli_2018}. Related to this covariant derivative, a  Levi-Civita--Carroll connection is introduced. The Levi-Civita--Carroll covariant derivative acts on a vector as
\begin{align}
   \hat{\nabla}_A X^B = \hat{\partial}_A X^B + \hat{\gamma}^B_{AC} X^C \, ,
\end{align}
where $\hat{\partial}_A = \partial_A + \frac{b_A}{\alpha} \partial_t$, and $\hat{\gamma}^B_{AC}$ is the Levi-Civita--Carroll connection \cite{Ciambelli_2018}. 

The commutation of the Levi-Civita--Carroll covariant derivative gives rise to the Riemann--Carroll tensor, which is covariant under Carrollian diffeomorphisms \cite{Ciambelli_2018}:
\begin{align}
   \left[ \hat{\nabla}_A, \hat{\nabla}_B \right] X^C = \hat{R}^C_{\hspace{4pt}DAB} X^D + \omega_{AB} \frac{2}{\alpha}\partial_t X^C \, .
\end{align}
Here, $\hat{R}^A_{\hspace{4pt}BCD}$ is the Riemann--Carroll tensor and $\omega_{AB}$ is the  vorticity.
Contracting the Riemann--Carroll tensor, the Ricci--Carroll tensor, $\hat{R}_{AB}$, is obtained:
\begin{align}
    \hat{R}^C_{\hspace{4pt}ACB} = \hat{R}_{AB} \, .
\end{align}
The Ricci--Carroll tensor can be expressed in terms of the geometric quantities defined on the horizon as follows \cite{freidel2024geometrycarrollianstretchedhorizons}: 
\begin{align}
    \hat{R}_{ AB } = \frac{1}{2} \hat{R} q_{AB} + \sigma_A^{\hphantom{A}C} \, \omega_{CB} + \frac{1}{2} \Theta \omega_{AB} \, ,
\end{align}
where $\hat{R}= q^{AB}\hat{R}_{AB}$ is its trace.
Therefore, its symmetric, traceless, and antisymmetric parts are given simply by
\begin{align}
    \hat{R}_{\langle AB \rangle} = \sigma_A^{\hphantom{A}C} \omega_{CB} \, , \quad \hat{R}_{\left[ AB \right]} = \frac{1}{2} \Theta \omega_{AB} \, .
\end{align}

Note that, within the gauge expressed in the main text, resulting in the metric in Eq.~(\ref{eq:metric}), the following choices are made: $\alpha=1$ and $b_A=0$. Therefore, the structures introduced in this appendix simplify accordingly. Specifically, the Riemann--Carroll and Ricci--Carroll tensors agree with the induced Riemann and Ricci tensors on the hypersurfaces. 
See the Appendix sections of \cite{Ciambelli_2018} and \cite{freidel2024geometrycarrollianstretchedhorizons} for more details on derivatives and connections summarized here.
\section{Gauge choices on the stretched Carrollian formalism \label{appendix:gaugechoices}}

Starting from the general metric on the bulk spacetime given in section 3 of \cite{freidel2024geometrycarrollianstretchedhorizons}, we imposed some conditions as discussed in the main text. Here, we summarize other possible choices and their implications for the horizon and Carrollian dynamics. 

\subsection{Rescaling the tangent vector $l^a$ \label{appendix:lgauge}}

The null tangent vector on the horizon can be rescaled while preserving all the properties of the Carrollian geometry. While we set the scaling parameter $\alpha$ to be $1$ in the main text, here we consider how this parameter affects the dynamics of the fluid and its stress-energy tensor.

We define a new tangent vector, which is related to $l^a$ as
\begin{align}
    \ell^a \equiv \alpha l^a \, .
\end{align}
We continue imposing the other coordinate choices in the main text.
Rescaling the tangent vector with a constant value does not alter the equations of motion. However, rescaling it with a function on the horizon, i.e., $\alpha = \alpha\left(v,\theta,\phi \right)$, induces a non-zero Carrollian acceleration and modifies the Hájíček one-form as follows:
\begin{align}
   \varphi_A &= - \frac{\partial_A \alpha }{\alpha} \\
    \mathcal{H}_A &= q_{AB}\partial_r V^B + \frac{\partial_A \alpha }{\alpha}
\end{align}
These, in turn, modify the DNS equation (Eq.~(\ref{eq:DNS})) to result in the following:
\begin{align}
    G_{lA} &= \partial_v{\mathcal{H}}_A  +{\Theta} \left( \mathcal{H}_A +\varphi_A \right)+ \notag\\ &\quad\left({\nabla}_B + \varphi_B \right)\left[ \sigma_A^{\hspace{4pt}B} - \left(\kappa +\frac{\Theta}{2}\right) \delta^B_A \right] \, . 
\end{align}
In the main text, setting $\alpha =1$ resulted in a vanishing acceleration on the horizon. This made it possible to close the system of equations with the four projections of the Einstein equations discussed in the main text. If the acceleration is non-zero, more relations are needed to solve the projected Einstein equations self-consistently.

The non-affinity parameter, $\kappa$, is also changed by the scaling of the tangent vector. The new value of $\kappa$ becomes
\begin{align}
    \kappa = \alpha \partial_r \rho + \partial_v \alpha \, .
\end{align}
From this, we see that we can choose the scaling function $\alpha$ in a way to ensure that the tangent vector is affinely parametrized, hence the non-affinity becomes $0$. In the cases, as in ours, where the stretching is only a function of $r$ and also in order $\mathcal{O}(r)$ (so that $\alpha$ is independent of $r$), we can integrate the equation $\kappa = 0$ to find the form of $\alpha$ that casts $l^a$ affine:
\begin{align}
    \alpha = e^{-(\partial_r \rho)v} \, .
\end{align}
For the perturbed Schwarzschild model we considered, this function becomes
\begin{align}
    \alpha = e^{-\frac{v}{4M}} .
\end{align}
This suggests that the null tangent $l^a$ should scale differently as the time evolves to keep the parametrization affine.

\subsection{Choices on $b_A$ and $V_A$ \label{appendix:bgauge}}

To construct the metric (Eq.~(\ref{eq:metric})), we used $b_A=0$ and $V_A\neq0$. Instead, if we had imposed $V_A=0$ and $b_A \neq 0$, then we would have a non-vanishing Carrollian acceleration $\varphi_A$, vorticity $\omega_{AB}$ on the horizon alongside a nonzero Hájíček field. The Carrollian acceleration explicitly appears in the horizon equations under this coordinate choice. Moreover, the covariant derivative in the expressions has to be changed with the Carroll covariant derivative that preserves Carrollian diffeomorphisms \cite{Ciambelli_2018} as summarized in Appendix~\ref{appendix:derivatives}. The associated partial derivative acts on a scalar function $f\left(x^i\right)$ on $\mathscr{H}$ as
\begin{align}
    \hat{\partial}_A f = \partial_A f + \frac{b_A}{\alpha} \partial_v f \;.
\end{align}
The resulting horizon equations become much more complex as the Levi-Civita--Carroll covariant derivatives and the Ricci--Carroll tensors in this case do not agree with the induced covariant derivative and the Ricci tensor on the hypersurface.

If instead, we follow the conventions of Randers--Papapetrou gauge in \cite{Donnay_2019, Ciambelli_2018, freidel2022carrollian} to have $b_A{\left(x^i\right)} \neq 0$ and $V_A=0$, while still keeping $\alpha=1$, we make the following intricate observation regarding the dynamics of the set of projected Einstein field equation on the horizon. We find that both $\varphi_A$ and $\mathcal{H}_A$ depend on $b_A$ on the horizon as
\begin{align}
    \varphi_A &\stackrel{\mathcal{N}}{=}  \partial_v b_A \\
    \mathcal{H}_A &\stackrel{\mathcal{N}}{=} - \frac{1}{2} b_A
\end{align}
Hence, the extra term of $\varphi_A$ appearing in the DNS (Eq.~(\ref{eq:DNS})) is not an independent function; rather, it satisfies $\varphi_A = -2 \mathcal{H}_A$. This, however, changes the character of the time evolution of the Hájíček field. The exponential behavior of the field converges to 0 as the null time evolves to infinity. This behavior is the opposite of that of the null expansion, which diverges in the infinite future as the Raychaudhuri equation (Eq.~
(\ref{eq:Raychaudhuri})) dictates. This well-known ``problem" in the horizon evolution is explained by its teleological nature \cite{TPM1986}.

The argument of teleology is briefly stated as follows: The event horizon is defined as a global null surface that is the boundary between null rays that can reach null infinity and those that cannot. This notion of the event horizon only makes sense from a global perspective where one knows the entire time evolution of the spacetime. Because of this, the horizon surface can behave acausally by responding to the outside fields before the fields reach the horizon itself. Such a consideration resolves the divergence in the Raychaudhuri equation and imposes correctly the boundary condition that the null expansion should vanish as the horizon evolves towards the infinite future. By the extension of this argument, we require that all our quantities behave similarly to the null expansions, i.e., diverge in the infinite future. Because of the observation that the  Hájíček field will converge in the infinite future, we were driven to impose the opposite condition where we set $b_A =0$ but allow instead $V_A \neq 0$. 

The teleological nature of the event horizon is also the reason why we implement a backward-in-time evolution to solve the horizon equations \cite{G_mez_2001}.

One could also impose a near-horizon peeling for the shift field by requiring that $b_A(x^a) = r \beta_A(x^i)$ (while $V^A(x^a) = 0$). This radial dependence guarantees that the Carrollian acceleration and vorticity both vanish on the horizon. However, the Hájíček field also vanishes. 

The general case as studied in \cite{freidel2024geometrycarrollianstretchedhorizons}  sets both $b_A \neq 0$ and $V_A\neq0$. Then the Hájíček field near the horizon behaves as
\begin{align}
    \mathcal{H}_A &\stackrel{\mathcal{N}}{=}  \frac{1}{2} \left(\partial_r V_A-  b_A\right)
\end{align}
This, however, introduces $b_A$ as an extra degree of freedom in the horizon equations. Therefore, more constraints need to be satisfied to solve the system of equations self-consistently. A set of such constraints called \textit{hydrodynamical horizon} is studied in \cite{freidel2024geometrycarrollianstretchedhorizons}. We intend to approach the resulting system with the numerical methods developed in this paper in future work.

\section{Horizon equations in the perturbative scheme \label{appendix:equationsO2}}

\textcolor{black}{
Here, we explicitly express the horizon equations expanded in the perturbative treatment up to $\mathcal{O}\left(\epsilon^2\right)$.
The equations in order $\mathcal{O}\left(1\right)$, that is, the order $\epsilon^0$, are identically satisfied by the background Schwarzchild metric.}

\begin{widetext}

\begin{itemize}
\item Order $\mathcal{O}(1):$
\begin{align}
  \hat{R}^{(0)} = \frac{1}{2M^2} \\[5pt]
  \hat{R}_{\langle AB \rangle}^{(0)} = 0 
\end{align}
    \item Order $\mathcal{O}(\epsilon)$:
\begin{align}
        \partial_v \Omega^{(1)}_{AB} &=  2M \partial_v \lambda_{AB}^{(1)}  + \frac{1}{2}\lambda_{AB}^{(1)}  -\frac{1}{2M} \Omega_{AB}^{(1)} -2M {\nabla}_{\langle A} \mathcal{H}_{B\rangle}^{(1)} \\[5pt]
        \partial_v \mathcal{H}_A^{(1)} &= {\nabla}_A \kappa^{(1)} - \frac{1}{2} \Omega^{\text{(Sph)}BC} {\nabla}_B \partial_v \Omega_{AC}^{(1)}\\[5pt]
        \kappa^{(1)}&= -M{\nabla}_A \mathcal{H}^{A(1)} + \frac{M}{2} \hat{R}^{(1)}
    \end{align}
  
    \item Order $\mathcal{O}(\epsilon^2)$:
\begin{align}
    &\partial_v \Theta^{(2)} = \frac{1}{4M} \Theta^{(2)} - \frac{1}{4} \partial_v \Omega^{AB (1)}\partial_v \Omega_{AB}^{(1)}  - \left( \partial_v \Phi^{(1)} \right)^2 \\[5pt]
    &\partial_v \Omega_{AB}^{(2)} =  2M \partial_v \lambda_{AB}^{(2)} + \frac{1}{2}\lambda_{AB}^{(2)} -\frac{1}{2M} \Omega_{AB}^{(2)} + 2\kappa^{(1)} \left(M \lambda_{AB}^{(1)} -  \Omega_{AB}^{(1)} \right) \notag \\ 
    &\qquad +\Omega^{\text{(Sph)}}_{AB} \left[M \partial_v \left(\Omega^{CD(1)}\right)\lambda^{(1)}_{CD} + 2M \Omega^{CD(1)}\partial_v\lambda_{CD}^{(1)}-\partial_v \left(\Omega^{CD(1)} \Omega_{CD}^{(1)}\right) \right.  \left. + \frac{1}{2} \Omega^{CD(1)}\lambda_{CD}^{(1)} - \frac{1}{2M}   \left(\Omega^{CD(1)} \Omega_{CD}^{(1)}\right) \right] 
   \notag\\
    & \qquad+4M \left( {\nabla}_{\langle A} \mathcal{H}_{B\rangle}^{(2)}  +  \mathcal{H}^{(1)}_{\langle A}  \mathcal{H}^{(1)}_{B\rangle} \right)  + 2M {\partial}_A \Phi^{(1)} {\partial}_B \Phi^{(1)} - M \Omega^{\text{(Sph)}}_{AB} \left( \Omega^{CD}_{\text{(Sph)}} {\partial}_C \Phi^{(1)} {\partial}_D \Phi^{(1)} \right) \\[5pt] 
    &\partial_v \mathcal{H}_A^{(2)} = {\nabla}_A \kappa^{(2)}
    -\frac{1}{2}{\nabla}_B \left[\Omega^{BC}_{\text{(Sph)}}\partial_v \Omega_{AC}^{(2)}+ \delta_A^{\hspace{4pt}B} \left( \Omega^{DE(1)}\partial_v \Omega_{DE}^{(1)}\right)\right]   + \frac{1}{2}\partial_A\Theta^{(2)} +  \partial_v \Phi^{(1)} \partial_A \Phi^{(1)} \\[5pt]
    &\kappa^{(2)} = -\frac{1}{2} \left(\partial_v \Omega^{AB(1)} \right) \left[\Omega_{AB}^{(1)} - M \lambda_{AB}^{(1)}\right] + \frac{M}{2} \Omega^{AB(1)} \partial_v \lambda_{AB}^{(1)} - \frac{1}{8M} \Omega^{AB(1)} \left[ \Omega_{AB}^{(1)} + M \lambda_{AB}^{(1)}\right] \notag\\
    & \qquad- M \left({\nabla}_A \mathcal{H}^{A(2)} + \mathcal{H}^{(1)}_A \mathcal{H}^{A(1)} \right)  + \frac{M}{2} \hat{R}^{(2)} - M \Omega^{AB}_{\text{(Sph)}} \partial_A \Phi^{(1)} \partial_B \Phi^{(1)} 
\end{align}
\end{itemize} 
\end{widetext}
 
\section{Details of numerical implementation \label{appendix:numerical}}

Starting from the spin-weighted functions introduced in Sec.~\ref {secV:level2},  we implement a spectral method to solve the system of coupled differential equations we obtained.

First, we decompose every scalar, vector, and tensor quantity into time-dependent coefficients multiplied by spherical harmonics of spin weights 0, 1, and 2, respectively. Below, we explicitly show the decompositions of the variables appearing in order $\mathcal{O}\left(\epsilon\right)$: 
\begin{align}
    k(v, x^A) &= \sum_{\ell, m} k_{\ell m}\left(v \right)  \,{}_0Y_{\ell m} \left(x^A \right) \, , \\
    h(v, x^A) &= \sum_{\ell , m} h_{\ell m}\left(v \right)  \,{}_1Y_{\ell m} \left(x^A \right) \, ,  \\
    c(v, x^A) &= \sum_{\ell , m} c_{\ell m}\left(v \right)  \,{}_2Y_{\ell m} \left(x^A \right) \, ,  \\
    s(v, x^A) &= \sum_{\ell , m} s_{\ell m}\left(v \right)  \,{}_2Y_{\ell m} \left(x^A \right) \, .
\end{align}
Similar decomposition is applied to the variables of $\mathcal{O}\left(\epsilon^2\right)$.

For the scalar field source, we follow the same expansion using spherical harmonics,
\begin{align}
    \varphi\left(v, x^A\right) &= \sum_{\ell , m} \varphi_{\ell m}\left(v \right)  \,{}_0Y_{\ell m} \left(x^A \right) \, .
\end{align}

After decomposing each term in the equations, we multiply both sides of the equations by the suitable spin-weighted spherical harmonic and integrate over the unit sphere to utilize the orthogonality property
\begin{align}
    \int_{S^1} {}_s\bar{Y}_{\ell m} \, {}_s{Y}_{\ell 'm'} = \delta_{\ell l'} \delta_{mm'} \, ,
\end{align}
where the bar denotes the complex conjugate and $\delta_{nn'}$ denotes the Kronecker delta. However, for the nonlinear terms appearing in order $\mathcal{O}\left(\epsilon^2\right)$, we need to incorporate the following integrals:
\begin{align}
    \int_\Omega {}_{s_1}\bar{Y}_{\ell_1m_1} \, {}_{s_2}{Y}_{\ell_2 m_2} \, {}_{s_3}{Y}_{\ell_3 m_3} = {}_{\left(s_1,s_2,s_3\right)}\mathcal{I}^{\ell_2m_2,\ell_3m_3}_{\ell_1m_1} \, .
\end{align}
In the case for which $-s_1 + s_2 + s_3 = 0$, these integrals can be computed via Wigner-$3j$ symbols as in \cite{Redondo_Yuste_2023}:
\begin{widetext}
\begin{align}
{}_{\left(s_1,s_2,s_3\right)}\mathcal{I}^{\ell_2m_2,\ell_3m_3}_{\ell_1m_1} = \sqrt{\frac{\left(2\ell_1 + 1\right)\left(2\ell_2 + 1\right)\left(2\ell_3 + 1\right)}{4\pi}} 
    \begin{pmatrix}
        \ell_1 & \ell_2 & \ell_3 \\
        m_1 & m_2 & m_3
    \end{pmatrix}
        \begin{pmatrix}
        \ell_1 & \ell_2 & \ell_3 \\
        s_1 & -s_2 & -s_3
    \end{pmatrix} \, .
\end{align}
\end{widetext}
Notably, for the expression we obtain, the condition  $-s_1 + s_2 + s_3 = 0$ is satisfied, and the numerical integration of the harmonics can be avoided using the $3j$-symbols in SymPy \cite{10.7717/peerj-cs.103}.

We further simplify the set of equations on the sphere by restricting ourselves to the case in which the effect of the bulk physics is incorporated only through the matter fields by imposing $\partial_rq_{AB} = \lambda = 0$, that is $s=0$, $S=0$. Moreover, as effects of the matter field only appear in order $\mathcal{O}(\epsilon^2)$, we tentatively set all the remaining first-order quantities $c$, $h$, $k$ to 0. Thus, we end up with the following system of equations describing the different modes of $\mathcal{O}(\epsilon^2)$ quantities:

\begin{widetext}
\begin{align}
    \partial_v \vartheta_{LM} &= \frac{1}{4M} \vartheta_{LM}  - \sum_{\ell_1,m_1} \sum_{\ell_2,m_2} {}_{\left(0,0,0\right)}\mathcal{I}^{\ell_1m_1,\ell_2m_2}_{LM} \partial_v \varphi_{\ell_1 m_1} \partial_v \varphi_{\ell_2 m_2} \\[5pt]
    \partial_v C_{LM} &=- \frac{1}{2M} C_{LM}   + 4M \sqrt{(L+1)(L-1)} H_{LM} \notag  \\ &\quad + 2M \sum_{\ell_1,m_1} \sum_{\ell_2,m_2}{}_{(-2,1,1)}\mathcal{I}_{LM}^{\ell_1 m_1, \ell_2 m_2} \sqrt{\ell_ 1 (\ell_1+1)} \sqrt{\ell_ 2 (\ell_2+1)}\varphi_{\ell_1 m_1}\varphi_{\ell_2 m_2}  \\
    \partial_v H_{LM} &= \sqrt{L (L+1)} K_{LM} + \frac{1}{8M^2} \sqrt{(L+2) (L-1)} \partial_v C_{LM} 
     \notag \\ & \quad + \frac{1}{2} \sqrt{L (L+1)} \vartheta_{LM} +\sum_{\ell_1,m_1} \sum_{\ell_2,m_2} {}_{(-1,0,1)}\mathcal{I}_{LM}^{\ell_1 m_1, \ell_2 m_2} \partial_v \varphi_{\ell_1 m_1} \sqrt{\ell_2 (\ell_2+1)} \varphi_{\ell_2 m_2} \\[5pt]
    K_{LM} &=  \frac{1}{8M} \sqrt{(L+1)L} \left[ H_{LM} - (-1)^{1-M}   \bar{H}_{L (-M)} \right]  + \frac{1}{64M^3} \sqrt{(L+2)(L+1)L(L-1)} \left[ C_{LM} + (-1)^{2-M}   \bar{C}_{L(-M)}  \right] \notag \\ & \quad + \frac{1}{4M} \sum_{\ell_1,m_1} \sum_{\ell_2,m_2}{}_{(0,1,-1)}\mathcal{I}_{LM}^{\ell_1 m_1, \ell_2 m_2}   \sqrt{\ell_1 (\ell_1+1)} \sqrt{\ell_2 (\ell_2+1)} \varphi_{\ell_1 m_1}   \varphi_{\ell_1 m_1} 
\end{align}
\end{widetext}

\subsection{Scalar field source profile \label{appendix:scalarfield}}
For each source mode $\varphi_{\ell m}$, we impose a Gaussian-like behavior with a growth phase until a time $v_0$ and a ringdown phase following the peak of the Gaussian. 
\begin{align}
    \varphi_{\ell m}\left(v \leq v_0\right) &=  \mathcal{A}_{\ell m} \exp{\left(-\frac{\left(v - v_0\right)^2}{2 \sigma^2}\right)} \cos{\left(f_{\ell m} \left(v - v_0\right)\right)} \notag \\
    \varphi_{\ell m}\left(v \geq v_0\right) &=  \mathcal{B}_{\ell m} \exp{\left( - \tau_{\ell m} \left(v - v_0 \right) \right)} \cos{\left(f_{\ell m} \left(v - v_0\right)\right)} \notag
\end{align}
where $\mathcal{A}_{\ell m}$ and $\mathcal{B}_{\ell m}$ are amplitudes, $\sigma$ is the width of the Gaussian, $\omega_{\ell m} = f_{\ell m} + i\tau_{\ell m} $ is the complex quasinormal mode (QNM) frequency and $v_0$ is the starting time of ringdown.

In the perturbed Schwarzschild model we study, we use Schwarzschild QNM frequencies for scalar modes \cite{Berti_2009, Stein:2019mop}. Specifically, we used the source modes $(\ell,m) = (0,0)$, $ (1,1)$, $ (1,-1)$, and $(2,0)$ as illustrated in Fig.~\ref{fig:source} in the main text. For simplicity, we set $v_0 = 0$ and $\sigma = 1$.

We chose the mode $(0,0)$ have the strongest amplitude, $\mathcal{A}_{00} = 1$, while we set $\mathcal{A}_{11}  = 0.5$, and $\mathcal{A}_{20}  = 0.25$. To make sure that the scalar field $\varphi$ is real-valued, we set $\varphi_{1-1}  = -\varphi_{11}^*$, and its amplitude accordingly.

The modes $\varphi_{1 \pm 1}$ break the axisymmetry in the perturbed spacetime. We keep these modes non-zero to observe the effects related to horizon angular momentum and a non-vanishing Hájíček field as discussed in the main text.

\section{Geometric quantities in example spacetimes \label{appendix:examplegauge}}

Here, we explicitly write the geometric quantities defined in Sec.\ref{secII:level1} for the Schwarzschild and slowly-rotating Kerr spacetimes. Our goal is to elucidate the consequences of our gauge choices in familiar spacetimes, as these are easier to interpret. Note that, in this appendix, we work with vacuum spacetimes; however, it is straightforward to extend the analysis to non-vacuum cases via a perturbation scheme as implemented in the main text.
\subsection{Schwarzschild}
The Schwarzschild metric in Eddington--Finkelstein coordinates with advanced null time is given by 
\begin{align}
    \mathrm{d} s^2 = - \left(1- \frac{2M}{r} \right) \mathrm{d} v^2 + 2 \mathrm{d}v \, \mathrm{d}r + r^2 \mathrm{d}\Omega^{2}_\mathrm{(Sph)} \, ,
\end{align}
where $\mathrm{d}\Omega^{2}_\mathrm{(Sph)} \equiv \mathrm{d}\theta^2 + \sin^2{\theta} \mathrm{d}\phi^2$.

From this metric, the following vectors and forms are identified to describe the horizon geometry as well as the rigging structure:
\begin{align*}
     l^a \partial_a &=   \partial_v  , &k^a \partial_a &= \partial_r , \\
       n_a \mathrm{d}x^a &=  \mathrm{d}r\quad  &\left(e_A\right)^a \partial_a &=   \delta^B_A \partial_B.
\end{align*}
The metric duals of these are computed using the metric and its inverse.
\begin{align*}
      l_a \mathrm{d}x^a &= -\left(1-\frac{2M}{r}\right) \mathrm{d}v +  \mathrm{d}r , & k_a \mathrm{d}x^a &= \mathrm{d}v, \\[5 pt]
       n^a \partial_a &=  \partial_v + \left(1-\frac{2M}{r}\right) \partial_r ,        &\left(e_{A}\right)_a \mathrm{d}x^a &=  q_{AB} \mathrm{d}x^B.
\end{align*}

Comparing these with the Carrollian metric in Eq.~(\ref{eq:metric}), we further identify
\begin{align}
    \rho = \frac{1}{2}\left(1 - \frac{2M}{r}\right) \, , && V_A =0 \, , && q_{AB} =  r^2 \Omega^{(\mathrm{Sph})}_{AB}\, .
\end{align}

Finally, we employ the near-horizon expansion. The event horizon of the Schwarzschild black hole is the null surface at $r=2M$. We define a new radial coordinate $\tilde{r} \equiv r-2M$ which vanishes on the horizon. We impose the near-horizon expansion by assuming $ 0 \leq \tilde{r} << 2M$ and expanding the expressions in orders of $\tilde{r}$. To compare with Eq.s~(\ref{eq:nearhorizonexp1}) \& (\ref{eq:nearhorizonexp2}), we expand the stretching and the metric induced on the base manifold up to $\mathcal{O}\left(\tilde{r}^2\right)$:
\begin{align}
    \rho&= \frac{1}{2}\left(1 - \frac{2M}{\tilde{r}+2M}\right) = \frac{\tilde{r}}{4M} + \mathcal{O}\left(\tilde{r}^2\right) \\[5pt]
    q_{AB} &=  \left(\tilde{r} - 2M\right)^2 \Omega^{(\mathrm{Sph})}_{AB}\\&= 4M^2 \Omega^{(\mathrm{Sph})}_{AB} - 4M\tilde{r} \Omega^{(\mathrm{Sph})}_{AB} + \mathcal{O}\left(\tilde{r}^2\right)
\end{align}
Then, the non-zero geometric quantities on the horizon are:
\begin{align}
    \kappa \stackrel{\mathscr{N}}{=} \frac{1}{4M} \, ,
    &&\bar{\Theta} \stackrel{\mathscr{N}}{=} \frac{1}{M} \, . 
\end{align}
Note that the components of the deviation tensor $B_{AB}$, defined with respect to the horizon tangent vector $l^a$, the Hájíček field, and the shear ($\bar{\sigma}_{AB}$) and vorticity ($\bar{\omega}_{AB}$) computed from the rigging vector $k^a$, all identically vanish on the horizon for the unperturbed Schwarzschild spacetime.

The Schwarzschild metric above is the unperturbed metric studied in Sec.\ref{secV:level1}.

\subsection{Slowly-rotating Kerr}
Another spacetime of interest here is the slow rotation limit of Kerr spacetime. We use this metric to gain insights about the Hájíček field and its relation to horizon angular momentum. If purely axisymmetric perturbations ($\varphi_{(\ell m)}$ with $m=0$) were applied to the background Schwarzchild metric in the perturbative scheme in Sec.~\ref{secV:level1}, the perturbed metric would have been studied with the slowly-rotating Kerr spacetime.

The slow rotation limit of Kerr spacetime is obtained from the standard Boyer–Lindquist coordinates by setting $a << 1$, where  $a \equiv J/M$ \cite{Poisson:2009pwt}. The line element in this limit becomes:
\begin{align}
    \mathrm{d}s^2_{\mathcal{M}} = -f \mathrm{d}t^2 + &\frac{1}{f} \mathrm{d}r^2 \notag \\&+ r^2 \mathrm{d}\Omega^2 - \frac{4Ma}{r} \sin^2{\theta} \mathrm{d}t \mathrm{d}\phi 
\end{align}
where $ f = 1 - 2M/r$.
This metric satisfies vacuum Einstein field equations up to $\mathcal{O}\left(a^2 \right)$. Therefore, the slow-rotation approximation introduces errors only at order $\mathcal{O}\left(a^2 \right)$. To work on this approximation systemically, we demand that all the equations are satisfied up to $\mathcal{O}\left(a^2 \right)$.

To express the slowly-rotating Kerr geometry with a null time coordinate and in the form similar to that of the Carrollian metric in Eq.~(\ref{eq:metric}), we implement the following coordinate transformations: 
\begin{align}
    v &\equiv t + r + 2M \ln{\left(\frac{r}{2M}-1\right)}  \\
    &\quad \Rightarrow
    \mathrm{d} v = \mathrm{d} t + \frac{1}{f} \mathrm{d} r \notag\\[5pt]
    \psi &\equiv \phi + \frac{a}{r} + \frac{a}{2M} \ln{\left(1 - \frac{2M}{r} \right)}\\
    &\quad \Rightarrow 
    \mathrm{d} \psi = \mathrm{d} \phi + \frac{2Ma}{r^3 f} \mathrm{d} r  \notag
\end{align}
These new coordinates are motivated in \cite{Poisson:2009pwt} as they are constant on the generators of null hypersurfaces defined by $v = \mathrm{constant}$. Thus, we obtained the following metric:
\begin{align}
    \mathrm{d}s^2 = - \left(1 - \frac{2M}{r} \right) \mathrm{d} v^2 + 2 \mathrm{d} v \mathrm{d} r + r^2 \mathrm{d} \Omega_{(\mathrm{Sph})}^2 \notag \\ - \frac{4Ma}{r} \sin^2{\theta} \mathrm{d}v \mathrm{d}\psi 
\end{align}
Here, we denote $\mathrm{d}\Omega^{2}_\mathrm{(Sph)} \equiv \mathrm{d}\theta^2 + \sin^2{\theta} \mathrm{d}\psi^2$. Now, we can identify the following vectors and forms as before:
\begin{align*}
     l^a \partial_a &=   \partial_v + \frac{2Ma}{r^3} \partial_\psi  , &k^a \partial_a &= \partial_r , \\
       n_a \mathrm{d}x^a &=  \mathrm{d}r\quad  &\left(e_A\right)^a \partial_a &=   \delta^B_A \partial_B.
\end{align*}
The metric duals of these are given by
\begin{align*}
      &l_a \mathrm{d}x^a = -\left(1-\frac{2M}{r}\right) \mathrm{d}v +  \mathrm{d}r , \\[5 pt]
      &k_a \mathrm{d}x^a = \mathrm{d}v, \\[5 pt]
       &n^a \partial_a =  \partial_v + \left(1-\frac{2M}{r}\right) \partial_r + \frac{2Ma}{r^3} \partial_\psi ,\\[5pt]
       &\left(e_{A}\right)_a \mathrm{d}x^a = - \frac{2Ma}{r} \sin^2{\theta} \mathrm{d}v +   q_{AB} \mathrm{d}x^B .
\end{align*}

Comparing these with the Carrollian metric in Eq.~(\ref{eq:metric}), we identify
\begin{align}
    \rho &= \frac{1}{2}\left(1 - \frac{2M}{r}\right) \, , \\[5pt]
    V^A &=   \frac{2Ma}{r^3}   \delta^{A}_{\psi} \, , \\[5pt]
    q_{AB} &=  r^2 \Omega^{(\mathrm{Sph})}_{AB}\, .
\end{align}

We further introduce a new radial coordinate $\tilde{r}$ that vanishes on the horizon $r= M +\sqrt{M^2 - a^2} =2M + \mathcal{O}\left(a^2\right)$. Note that the corrections to the location of the event horizon are of order $\mathcal{O}\left(a^2\right)$, hence are neglected in the slow rotation limit. As we did for the Schwarzschild case, we use this new radial coordinate to implement the near-horizon expansion, for which we  $ 0 \leq \tilde{r} << 2M$ and expand the expressions in orders of $\tilde{r}$. 
\begin{align}
    \rho &= \frac{1}{2}\left(1 - \frac{2M}{2M + \tilde{r}}\right) = \frac{\tilde{r}}{4M} + \mathcal{O}\left(\tilde{r}^2\right) \\[5pt]
    V^A &=   \frac{2Ma}{\left(2M + \tilde{r}\right)^3}  \delta^{{\psi}}_A \, , 
    \notag\\&=  \frac{a}{4M^2} \delta^{{\psi}}_A -  \frac{3a\tilde{r}}{8M^2}   \delta^{{\psi}}_A   + \mathcal{O}\left(\tilde{r}^2\right)\,    \\[5pt]
    q_{AB} &=  \left( 2M +  \tilde{r}\right)^2 \Omega^{(\mathrm{Sph})}_{AB} \notag\\&= 4M^2 \Omega^{(\mathrm{Sph})}_{AB} - 4M\tilde{r} \Omega^{(\mathrm{Sph})}_{AB} + \mathcal{O}\left(\tilde{r}^2\right)\, 
\end{align}

Then, the non-zero geometric quantities on the horizon are:
\begin{equation}
     \kappa \stackrel{\mathscr{N}}{=} \frac{1}{4M}\, ,
\end{equation}
\begin{equation}
     \bar{\Theta} \stackrel{\mathscr{N}}{=} \frac{1}{M}\, ,
\end{equation}
\begin{equation}
     \mathcal{H}_A \stackrel{\mathscr{N}}{=} -\frac{3a}{4M} \sin^2\theta \delta^{A}_{\psi} \, .
\end{equation}
All other geometric quantities introduced in Sec.~\ref{secII:level1} vanish on the horizon.

Finally, we compute the horizon angular momentum using the Hájíček field. As the spacetime is axisymmetric, there exists the axial Killing vector which can be expressed in these coordinates as $\partial_{{\psi}}$. Following Damour's derivation \cite{Damour1982}, we compute the angular momentum of the horizon via
\begin{align}
    J_{\mathcal{N}} &= - \frac{1}{8\pi} \int \mathcal{H}_{{\psi}} |_{\mathcal{N}} \, \mathrm{d}S_{\mathcal{N}}\\[5pt]
    &= -\frac{1}{8\pi} \int \left(-\frac{3Ma}{4M^2} \sin^2{\theta}\right) \, 4M^2\sin{\theta} \, \mathrm{d}\theta  \, \mathrm{d}{{\psi}} \\[5pt]
    &= Ma \, .
\end{align}
This is the expected value of the event horizon angular momentum for the slowly-rotating Kerr metric.

The identifications from the slowly-rotating Kerr spacetime to the geometric quantities that are useful for the Carrollian fluid/horizon dictionary can be extended to the full Kerr solution at the cost of trading the simple results of this appendix for much more complex relations.

\bibliography{apssamp}

\end{document}